\documentclass[reprint,nofootinbib,superscriptaddress,amsmath,amssymb,aps,prr,longbibliography]{revtex4-1}

\usepackage[T1]{fontenc}
\usepackage{txfonts}
\usepackage{bm,dsfont,mathtools}   % for math
\usepackage{graphicx}% Include figure files
\graphicspath{ {./figures/} }
\usepackage{dcolumn}% Align table columns on decimal point
\usepackage{physics}
\usepackage{siunitx}
\usepackage{xcolor}
\usepackage{lipsum}
\usepackage[colorlinks=true,breaklinks=true,allcolors=blue]{hyperref}
\usepackage[makeroom]{cancel}
\usepackage[normalem]{ulem}

\newcommand{\Fig}[1]{Fig.~\ref{fig:#1}}
\newcommand{\App}[1]{App.~\ref{app:#1}}

\newcommand{\e}{\mathrm{e}}
\newcommand{\pL}{p_{\mathrm{IR}}}
\renewcommand{\i}{\mathrm{i}}

\newcommand{\eq}[1]{(\ref{eq:#1})}
\newcommand{\Eq}[1]{Eq.\,\eq{#1}}

\newcommand{\fig}[1]{\ref{fig:#1}}

\newcommand{\Sect}[1]{Sec.~\ref{sec:#1}}

\makeatletter
\let\cat@comma@active\@empty
\makeatother

\AtBeginDocument{%
    \newwrite\bibnotes
    \def\bibnotesext{Notes.bib}
    \immediate\openout\bibnotes=\jobname\bibnotesext
    \immediate\write\bibnotes{@CONTROL{REVTEX41Control}}
    \immediate\write\bibnotes{@CONTROL{%
    apsrev41Control,author="08",editor="1",pages="1",title="0",year="1"}}
     \if@filesw
     \immediate\write\@auxout{\string\citation{apsrev41Control}}%
    \fi
}%

\begin{document}

%\preprint{APS/123-QED}

\title{Kelvin waves in nonequilibrium universal dynamics of relativistic scalar field theories}

\author{Viktoria Noel}
\affiliation{Institute for Theoretical Physics, 
        Ruprecht-Karls-Universit\"at Heidelberg, 
        Philosophenweg 16, 
        69120~Heidelberg, Germany}
\affiliation{Kirchhoff Institute for Physics,
		Ruprecht-Karls-Universit\"at Heidelberg,
		Im Neuenheimer Feld 227,
		69120~Heidelberg, Germany}
        
\author{Thomas Gasenzer}
\affiliation{Kirchhoff Institute for Physics,
		Ruprecht-Karls-Universit\"at Heidelberg,
		Im Neuenheimer Feld 227,
		69120~Heidelberg, Germany}
\affiliation{Institute for Theoretical Physics, 
        Ruprecht-Karls-Universit\"at Heidelberg, 
        Philosophenweg 16, 
        69120~Heidelberg, Germany}

\author{Kirill Boguslavski}
\affiliation{Institute for Theoretical Physics, 
        TU Wien, 
        Wiedner Hauptstraße 8-10, 
        1040 Vienna, Austria}
\date{\today}

\begin{abstract}
We investigate the different degrees of freedom underlying far-from-equilibrium scaling behaviour in a relativistic, single-component $\mathrm{O}(1)$ scalar field theory in two and three spatial dimensions.
In such a strongly correlated many-body system, identifying the respective roles of nonlinear wave excitations and defect dynamics is a prerequisite for understanding the universal character of time evolution far from equilibrium and thus the different possible universality classes of nonthermal fixed points.
Using unequal-time two-point correlation functions, we extract information about the dominant infrared excitations and study their connections to the turbulent dynamics of topological defects created in the system. 
In three dimensions, the primary excitations are identified as kelvon quasiparticles, which are quantised Kelvin waves propagating along vortex lines, while in two dimensions, vortices are point defects, and the infrared dynamics is dominated by bound-state like excitations similar to Kelvin waves.
In both cases, the kelvon excitations are found to be characterised by distinct time-evolving dispersion relations, subject to the coarsening dynamics close to the respective nonthermal fixed point and, thus, to the decay of superfluid turbulence in the system. 
Our results underline the role of topological defects and their influence on the universal dynamics of strongly correlated systems near nonthermal fixed points, complementing the analysis of large-$N$ models in $\mathrm{O}(N)$ systems.
\end{abstract}

\maketitle

%\tableofcontents
% ====================================================================
% ==================================================================== 
\section{Introduction}
\label{sec:introduction}
The concept of universality is a powerful framework for understanding complex many-body systems.
Near equilibrium, it provides a way to associate phase transitions with universality classes based on a small set of system parameters, highlighting how vastly different systems can exhibit similar dynamical critical behaviour \cite{Hohenberg:1977ym}.
Beyond this, universality can also emerge far from equilibrium.
Fluid turbulence represents one of the first scaling phenomena systematically discussed in physics \cite{Frisch1995a}.
Quantum coherence allows far-from-equilibrium scaling to emerge in low-energy as well as relativistic systems, in particular in the form of superfluid or quantum turbulence  
\cite{Svistunov2001a,
Volovik2004a,
Vinen2006a,
Tsubota2008a,
Proment2010a,
Micha:2002ey,
Henn2009a.PhysRevLett.103.045301,
Kwon2014a.PhysRevA.90.063627,
Johnstone2019a.Science.364.1267,
Glidden:2020qmu,
Mueller2021a.PhysRevX.11.011053,
Mueller2024.PhysRevLett.132.094002} 
and different types of wave turbulence
\cite{Zakharov1992a,
Nazarenko2011a,
kozik2004kelvin,
Mueller:2006up,
Lvov2010a,
Boue2011a.PhysRevB.84.064516,
Navon2016a.Nature.539.72,
Navon2018a.Science.366.382},
studied both in theory and experiment.
Turbulence is commonly considered in an open-system setting, where driving and dissipation give rise to a stationary cascade that exhibits spatial scale invariance within an inertial range of energy scales.

Universal dynamics far from equilibrium can also occur in closed systems exhibiting self-similar scaling in space and time. 
Such universality has been studied in analogy to (near-)equilibrium scaling, in connection with nonthermal fixed points \cite{Berges:2008wm, Berges:2008sr, Scheppach:2009wu, Nowak:2011sk, Schole:2012kt, PineiroOrioli:2015cpb, Walz:2017ffj, Chantesana:2018qsb, Boguslavski:2019ecc}, 
which give rise to nonequilibrium attractor solutions. 
This far-from-equilibrium universality has been experimentally observed in ultracold atomic gases in different dimensions and trapping geometries 
\cite{%
Henn2009a.PhysRevLett.103.045301,
Gring2011a,
AduSmith2013a,
Kwon2014a.PhysRevA.90.063627,
Langen2015b.Science348.207,
Navon2015a.Science.347.167N,
Navon2016a.Nature.539.72,
Rauer2017a.arXiv170508231R.Science360.307,
Gauthier2019a.Science.364.1264,
Johnstone2019a.Science.364.1267,
Eigen2018a.arXiv180509802E,
Prufer:2018hto,
Erne:2018gmz,
Navon2018a.Science.366.382,
Glidden:2020qmu,
GarciaOrozco2021a.PhysRevA.106.023314,
Lannig:2023fzf,
Martirosyan:2023mml,
Gazo:2023exc,
MorenoArmijos2024a.PhysRevLett.134.023401,
Martirosyan:2024rxm}.
Many phenomena and properties of universal scaling dynamics have been theoretically predicted and studied in very different systems, ranging from ultracold gases over heavy-ion collisions to the early universe,
\cite{%
Kodama:2004dk,
Mukerjee2007a.PhysRevB.76.104519,
Berges:2008wm, 
Berges:2008sr,
Scheppach:2009wu,
Barnett2011a,
Nowak:2011sk,
Nowak:2012gd,
Schole:2012kt,
Berges:2012us, 
Karl:2013kua,
Karl:2013mn,
Gasenzer:2013era,
Berges:2013lsa,
Marcuzzi2013a.PhysRevLett.111.197203,
Langen2013a.EPJST.217,
Berges:2014bba,
Ewerz:2014tua,
Berges:2015ixa,
Moore:2015adu,
Nessi2014a.PhysRevLett.113.210402,
Gagel2014a.PhysRevLett.113.220401,
Bertini2015a.PhysRevLett.115.180601,
Babadi2015a.PhysRevX.5.041005,
Buchhold2015a.PhysRevA.94.013601,
PineiroOrioli:2015cpb, 
Hofmann2014a,
Maraga2015AgingCoarsening,
Williamson2016a.PhysRevLett.116.025301,
Williamson2016a.PhysRevA.94.023608,
Villois2016a.PhysRevE.93.061103,
Bourges2016a.arXiv161108922B.PhysRevA.95.023616,
Chiocchetta:2016waa.PhysRevB.94.174301,
Karl:2016wko,
Berges:2017ldx,
Walz:2017ffj, 
Deng:2018xsk,
Mazeliauskas:2018yef, 
Baggaley2018a.PhysRevA.97.033601,
Bland2018a.PhysRevLett.121.174501,
Schmied:2018upn.PhysRevLett.122.170404,
Mikheev:2018adp,
Chantesana:2018qsb, 
Schlichting:2019abc,
Boguslavski:2019fsb, 
Berges:2019oun, 
Williamson2019a.ScPP7.29,
Schmied:2018osf.PhysRevA.99.033611,
Schmied:2019abm,
Spitz2021a.SciPostPhys11.3.060,
Gao2020a.PhysRevLett.124.040403,
Groszek2020a.SciPostPhys.8.3.039,
Groszek2021a.PhysRevResearch.3.013212,
Wheeler2021a.EPL135.30004,
Gresista:2021qqa,
Liu:2022rss,
Mikheev:2022fdl, 
Heinen:2022rew,
Heinen2023a.PhysRevA.107.043303,
Siovitz:2023ius.PhysRevLett.131.183402,
Berges:2023sbs, 
Spitz:2023wmn, 
Heller:2023mah, 
Noel2024:PhysRevD.109.056011,
Siovitz:2024aqi,
Spitz:2024jeg, 
Rosenhaus2024a.PhysRevE.109.064127,
Rosenhaus2024a.PhysRevLett.133.244002,
Rosenhaus:2024iqw,
Rosenhaus:2025mgj}, 
for overviews cf.~%
\cite{Nowak:2013juc,Berges:2015kfa,Schmied:2018mte,Berges:2020fwq,Siovitz:2023uwo}.
This includes studies of (wave) turbulence, as well as coarsening and phase-ordering kinetics \cite{Bray1994a.AdvPhys.43.357,Puri2019a.KineticsOfPT,Cugliandolo2014arXiv1412.0855C}.
The vast range of phenomena raises the question of what the relevant physics behind the apparent universality is, which universality classes exist and whether these classes can be divided into subclasses. 

The study of far-from-equilibrium universality in isolated systems has primarily focused on the properties of the time-dependent momentum distribution function, $f(t, p)$, which describes the occupancy of momentum modes over time. 
These functions reveal self-similar scaling behaviour near nonthermal fixed points, characterised by universal scaling exponents that are largely independent of the systems' parameters or initial conditions.  
Relativistic scalar field theories with $\mathrm{O}(N)$ symmetry,
for example, exhibit universal dynamics at low momenta according to $f(t, p)=t^{\,\alpha} f_\mathrm{s}(t^{\,\beta} p)$, with the exponents $\alpha = d\beta$ and  $\beta = 1/2$ \cite{PineiroOrioli:2015cpb}. 
The scaling function $f_\mathrm{s}$ is consistent across different values of $N$, whether in relativistic or nonrelativistic models, and the underlying physics is connected to the self-similar transport of particle numbers toward lower momenta and the growth of a quasi condensate in the long-wavelength modes below a characteristic scale $\pL(t)\sim t^{-\beta}$.

Large-$N$ kinetic theories provide a successful description of this scaling dynamics and the exponents, which can be traced to weak elastic collisions of Goldstone quasiparticles with free dispersion and momentum-dependent effective quartic coupling \cite{Berges:2008wm, Berges:2008sr, Scheppach:2009wu, Berges:2010ez, PineiroOrioli:2015cpb, Berges:2015ixa, Berges:2016nru, Walz:2017ffj, Chantesana:2018qsb, Mikheev:2018adp, Rosenhaus2024a.PhysRevLett.133.244002, Rosenhaus:2025mgj}, representing the excitations of the phase-angle differences of the different components \cite{Mikheev:2018adp, Boguslavski:2019ecc}. 
For small values of $N$, however, where the excitations of the total field amplitude become relevant, the precise description of the universality remains unsatisfactory, and alternative mechanisms, such as coarsening dynamics of topological defects are anticipated to play a significant role \cite{nowak2011superfluid, Gasenzer:2011by, Nowak:2011sk, Schole:2012kt, Gasenzer:2013era, Karl:2013kua, Ewerz:2014tua, Karl:2016wko, Deng:2018xsk, Johnstone2019a.Science.364.1267, Schmied:2018osf.PhysRevA.99.033611, Schmied:2019abm, Siovitz:2023ius.PhysRevLett.131.183402, Lannig:2023fzf, Huh:2023xso}. 
Recent studies indicate that for systems with a small number of field components, distinct initial conditions may lead to different scaling exponents \cite{Karl:2016wko, Lannig:2023fzf, Heinen2023a.PhysRevA.107.043303, Siovitz:2024aqi}, where distinctly sub-diffusive scaling, $\beta \ll 1/2$, has been associated with specific defect-driven dynamics, e.g., $\beta\approx1/5$ for vortices in two spatial dimensions \cite{Karl:2016wko,Johnstone2019a.Science.364.1267,Spitz2021a.SciPostPhys11.3.060,Noel2024:PhysRevD.109.056011}.

For the relativistic $\mathrm{O}(N)$-symmetric nonlinear Klein-Gordon vector models in both two and three spatial dimensions, a number of different, $N$-dependent topological defects are expected to contribute to the dynamics in the infrared~\cite{Moore:2015adu, Deng:2018xsk, Noel2024:PhysRevD.109.056011}, which includes vortex defects in $\mathrm{O}(1)$.
The topological connection between the $\mathrm{O}(1)$ and the nonrelativistic $\mathrm{U}(1)$ theory, and hence, the existence and definition of vortices in the relativistic $\mathrm{O}(1)$ model originates from an emerging mass gap. 
Therefore, at low energies, the field and its canonical conjugate are mapped to the real and imaginary parts of the complex $\mathrm{U}(1)$ field.
However, for $\mathrm{O}(N)$ theories in three (spatial) dimensions, including the case $\mathrm{O}(1)$, the self-similar scaling evolution of the distribution function has been consistent with that of large-$N$ descriptions and in agreement with nonrelativistic $\mathrm{U}(N)$-symmetric nonlinear Schr\"odinger models ~\cite{Scheppach:2009wu, PineiroOrioli:2015cpb, Moore:2015adu, Walz:2017ffj, Chantesana:2018qsb, Mikheev:2018adp, PineiroOrioli:2018hst, Boguslavski:2019ecc}. 
This has been corroborated by observing the scaling behaviour of $f(t,p)$ based on classical-statistical (Truncated-Wigner) real-time lattice simulations. 
Nonetheless, unequal-time correlation functions, which also give information about the nonequilibrium excitation spectrum, have revealed markedly different dominant excitation peaks for $N=1,2$ compared to $N \geq 3$, where the large-$N$ excitation indeed dominates \cite{Boguslavski:2019ecc}. 

Here we study the underlying mechanisms of the self-similar infrared transport for the relativistic single-component $\mathrm{O}(1)$ theory in two and three spatial dimensions using unequal-time two-point correlation functions. 
By providing information on both occupancies and dispersion relations, the latter observables allow us to distinguish the different excitations contributing to the scaling and also to ascertain which excitation is dominant. 
Based on numerical simulations in three spatial dimensions, the dominant excitation in such correlators at large $N$ has been identified as the one with the (complex) quasiparticle dispersion \cite{Shen:2019jhl, Boguslavski:2019ecc} that agrees with the kinetic theory picture.
An excitation similar to the large-$N$ excitation has also been found for the relativistic $N=2$ case at low momenta \cite{Boguslavski:2019ecc}, whereas it is absent for $N=1$.
On the other hand, a very specific, so-called ``transport peak'' has been identified for both nonrelativistic $\mathrm{U}(1)$ and in the nonrelativistic limit of relativistic $\mathrm{O}(1)$ models \cite{PineiroOrioli:2018hst, Boguslavski:2019ecc}, with the same momentum and time dependence of the dispersion relation, as well as damping rate, while being distinct from the large-$N$ excitation. 
This peak appears to be present for the cases of $\mathrm{O}(2)$ and $\mathrm{O}(3)$ as well, however, it distinctly dominates only the $N=1$ theory \cite{Boguslavski:2019ecc}. 
The origin of this transport peak has not been understood yet and is not predicted by any current kinetic or low-energy effective description of nonthermal fixed points. 

In anticipation of its potential connection to topological defects, in this work, we also extract and analyse the vortex dynamics for the $\mathrm{O}(1)$ theory in both two and three dimensions using topological defect data extracted from lattice configurations.
In the context of nonthermal fixed points, two-dimensional relativistic theories have not been extensively studied yet \cite{Deng:2018xsk, Gasenzer:2011by, Gasenzer:2013era}. 
However, they do provide an interesting toy model since, intuitively speaking, topological effects are more important in lower dimensions and for lower $N$.  
By carefully investigating the transport peak, we find a direct connection between the self-similar infrared transport near the nonthermal fixed point and vortex line defects in three dimensions in terms of Kelvin waves \cite{Thomson1880, Pitaevskii1961a, Gross:1961, Donnelly2005a}, which arise as quantised excitations on vortices present in scalar theories, and which are well-investigated in the context of vortex dynamics and quantum turbulence \cite{vinen2001decay,vinen2003kelvin,kozik2004kelvin,Kozik2005a.PhysRevLett.94.025301,kozik2005vortex,Kozik2008a.PhysRevLett.100.195302,Kozik2009a, Lvov2010a, Krstulovic2012a.PhysRevE.86.055301,Laurie2010a.PhysRevB.81.104526, Boue2011a.PhysRevB.84.064516, boue2015energy}. 
In two dimensions, we observe a similar transport peak, where the underlying physics is connected to the two-dimensional analogues of Kelvin waves, which arise as vortex-displacement excitations \cite{Isoshima1999a.PhysRevA.59.2203, simula2013collective, simula2018vortex}.
While the general picture of nonlinear wave propagation in the strongly correlated regime near a nonthermal fixed point still applies for both small and large values of $N$, with the present study, we distinctly identify the relevant degrees of freedom in these theories, highlighting the importance of topological defects for a single-component scalar field theory. 

The paper is organised as follows: \Sect{model} describes the lattice simulations and the different types of two-point correlation functions investigated. 
\Sect{dof} provides an overview of the relevant degrees of freedom based on applicable low-energy effective theories of relativistic scalar fields. 
It discusses the expected $N$-dependence from both the quasiparticle picture and the possible topological defects. 
In \Sect{results}, we present our main findings from real-time lattice simulations. 
We first compare the self-similar scaling from the (equal-time) distribution function to that of vortex-antivortex annihilation-driven coarsening dynamics and then draw parallels to the relevant excitations found in unequal-time correlation functions.
We analyse the dispersion relation of these excitations to clearly pinpoint their nature.
In Section~\ref{sec:discussion}, we present our conclusions.

% ==================================================================== 
% ==================================================================== 
\section{Model and observables}
\label{sec:model}

We consider a single-component relativistic 
scalar field theory with field variable $\phi(t,\bold{x})$, in $d=2$ and $d=3$ spatial dimensions with classical action
\begin{equation}
		S[\phi]=\int_{t, \bold{x}}\left[\frac{1}{2} \partial^\mu \phi \partial_\mu \phi-\frac{m^2}{2} \phi^2-\frac{\lambda}{4 !}\phi^4\right]\,,
        \label{eq:OoneModel}
\end{equation}
where $\int_{t, \bold{x}} \equiv \int \mathrm{d} t \int \mathrm{d}^d x$, summation over repeated indices is implied, $m$ is the bare mass and $\lambda$ is the coupling constant. 
In the weak-coupling regime $\lambda \ll 1$ considered in this work, the quantum dynamics can be accurately mapped onto classical-statistical field theory (Truncated-Wigner approximation) for highly occupied systems at not too late times \cite{Polkovnikov:2009ys}. 
In classical-statistical simulations, one samples the fields over initial conditions,
\begin{equation}
\begin{aligned}
& \phi(0, \mathbf{p})=\sqrt{f(0, p) / p}\, c(\mathbf{p})\,, \\
& \pi(0, \mathbf{p})=\sqrt{p f(0, p)}\, \tilde{c}(\mathbf{p})\,,
\end{aligned}
\end{equation}
where $p = |\mathbf{p}|$, with independent random numbers drawn from a Gaussian distribution that satisfy
$\left\langle c(\mathbf{p})\left(c(\mathbf{q})\right)^*\right\rangle_{\mathrm{cl}}=V \delta_{\mathbf{p}, \mathbf{q}}$, and similarly for $\tilde{c}$.
Our initial condition sets high occupations for the infrared modes, 
\begin{equation}
\label{eq:boxinicond}
f(t=0, \mathbf{p})=\frac{n_0}{\lambda} \Theta(Q-p)\,,
\end{equation}
up to a characteristic momentum scale $Q$, which is the physical scale governing the dynamics. 
Subsequently, each realisation is evolved according to the classical equations of motion,
\begin{equation}
\begin{aligned}\label{eq:class_EOM}
& \partial_t \phi(t, \mathbf{x})=\pi(t, \mathbf{x})\,, \\
& \partial_t \pi(t, \mathbf{x})=\partial_i \partial_i \phi(t, \mathbf{x})-m^2 \phi(t, \mathbf{x})-\frac{\lambda}{6} \phi(t, \mathbf{x})^3.
\end{aligned}
\end{equation}
In the numerical approach, we discretise $\phi(t,\mathbf{x})$ and $\pi(t,\mathbf{x})$ on a spatial lattice with spacing $a_s$, volume $V=(N_sa_s)^d$ and time step $a_t=0.05\,a_s$. 
In $d=2$ dimensions, we use $N_s=2048$, $a_s=0.1\,Q^{-1}$, $m/Q=2$, $n_0=80$. 
In $d=3$ dimensions, we use $N_s=256$, $a_s=0.8\,Q^{-1}$, $m/Q=0.25$, $n_0=100$, unless stated otherwise.
We use a fourth-order accurate finite-difference scheme for the second-order spatial derivatives as described in \cite{Micha:2002ey}. The time evolution is performed using a leapfrog algorithm in propagating Eqs.~\eqref{eq:class_EOM}, which conserves the energy density and where the fields $\phi$ and $\pi$ are half a timestep apart. 
We furthermore impose periodic boundary conditions for the spatial lattice.

Any dependence on the coupling constant drops out after the classical equations of motion and the initial conditions are rescaled according to $\phi \rightarrow \lambda^{1 / 2} \phi$, $\pi \rightarrow \lambda^{1 / 2} \pi$. This property stems from the underlying classical limit $\hbar \to 0$ or, equivalently, the weak coupling limit $\lambda \rightarrow 0$ while $\lambda f$ remains unchanged.
The figures shown in the following sections all assume quantities in units of $Q$.

To study the excitation spectrum, we will compute two different unequal-time two-point correlation functions: the spectral function $\rho(t^{\prime}, t, \mathbf{p})$ and the statistical correlation function $F(t^{\prime}, t, \mathbf{p})$. 
The latter can be defined as the expectation value of an anticommutator of fields, which can be calculated in the classical-statistical framework as the classical expectation value of the corresponding product of fields,
\begin{equation}
F\left(t^{\prime}, t, \mathbf{p}\right)=\frac{1}{V}\left\langle \phi\left(t^{\prime},\mathbf{p}\right) \, \phi^*(t, \mathbf{p}) \right\rangle_{\mathrm{cl}}.
\end{equation}
Based on this, we can also define the distribution function as 
\begin{equation}
\label{eq:Defftp}
f(t, \mathbf{p})=\left.\sqrt{F\left(t^{\prime}, t, \mathbf{p}\right) \partial_t^{\prime} \partial_t F\left(t^{\prime}, t, \mathbf{p}\right)}\right|_{t'=t}
\end{equation}
at equal times $t'=t$.

For the computation of the spectral function, the linear-response approach of \cite{Boguslavski:2019ecc} similar to \cite{Boguslavski:2018beu, PineiroOrioli:2018hst} is employed. 
This involves perturbing the system, at time $t$, as $\phi \rightarrow \phi+\delta \phi$ with a source term 
$j(t, \mathbf{x})=j^0(\mathbf{x}) \delta\left(t-t^{\prime}\right)$, drawn from a random Gaussian distribution that satisfies $\left\langle j^0(\mathbf{p})\left(j^0(\mathbf{q})\right)^*\right\rangle_j=V \delta_{\mathbf{p}, \mathbf{q}}$. 
The perturbation $\delta \phi$ is time evolved according to the linearised equations of motion $\partial_{t}^{\prime} \delta \phi=\delta \pi$ and
\begin{equation}
\begin{aligned}
\partial_{t}^{\prime} \delta \pi=\partial_i \partial_i \delta \phi-\left(m^2+\frac{\lambda}{6} \phi^2\right) \delta \phi - j^0\, \delta\left(t-t^{\prime}\right)\,.
\end{aligned}
\end{equation}
The retarded component of the spectral function follows from linear response theory as
\begin{equation}
\theta\left(t^{\prime} - t \right) \rho\left(t^{\prime}, t, \mathbf{p}\right)=\frac{1}{V} \left\langle \left\langle \delta\phi\left(t^{\prime}, \mathbf{p}\right) \right\rangle_{\mathrm{cl}} \left(j^0(\mathbf{p})\right)^* \right\rangle_j.
\end{equation}
Apart from the occupancy of excitations in the system, the correlators $F$ and $\rho$ also contain information about the frequency dependence of excitations, including the dispersion relations and damping rates of all of the relevant excitation species in the system.
We consider these unequal-time correlation functions in momentum space and perform the Fourier transforms with respect to relative time $\Delta t=t^{\prime}-t''$ as
\begin{align}
\label{eq:FTDefFtauomegap}
F(t, \omega, \mathbf{p}) 
&\equiv \int \mathrm{d}\Delta t\, \e^{\i \omega \Delta t} F\left(t^{\prime}, t'', \mathbf{p}\right)\,,
\\
\label{eq:FTDefrhotauomegap}
\rho(t, \omega, \mathbf{p}) 
&\equiv-\i \int \mathrm{d}\Delta t \,\e^{\i \omega \Delta t} \rho\left(t^{\prime}, t'', \mathbf{p}\right)\,,
\end{align}
where $t \equiv\left(t^{\prime}+t''\right) / 2$ is the (central) time. 
In practice, we evaluate the correlations at $t'' = t$.
This approximation can be justified provided the correlations do not change significantly over times of the order of their oscillation periods along the relative time $\Delta t$, 
which is fulfilled during the self-similar scaling dynamics \cite{PineiroOrioli:2018hst}. 
Therefore, in our simulations, we compute the Fourier transforms while holding $t$ fixed, as
\begin{align}
F(t, \omega, \mathbf{p}) 
&\approx 2 \int_0^{\Delta t_{\max }}\!\! \mathrm{d}\Delta t \,\cos (\omega \Delta t)\, h(\Delta t)\, F\left(t+\Delta t, t, \mathbf{p}\right),
\\
\rho(t, \omega, \mathbf{p}) 
&\approx 2 \int_0^{\Delta t_{\max }}\!\! \mathrm{d}\Delta t\, \sin (\omega \Delta t)\, h(\Delta t)\, \rho\left(t+\Delta t, t, \mathbf{p}\right),
\end{align}
typically using $\Delta t_{\mathrm{max}}\approx300$ for spectral functions and $\Delta t_{\mathrm{max}}\approx500$ for the statistical function.
We have checked that the results are not sensitive to the exact value of $\Delta t_{\mathrm{max}}$.
We also employ a Hann window function, 
\begin{equation}
h(\Delta t)=\frac{1}{2}\left(1+\cos \frac{\pi \Delta t}{\Delta t_{\max }}\right),
\end{equation}
and zero padding to smoothen the resulting curves, and have checked that these do not alter the form of the displayed peaks. 
Expectation values of observables are obtained by averaging over the field trajectories that are solutions to the classical equations of motion. 
The results shown in the following use correlation functions averaged over 100 realisations in two and 40 realisations in three dimensions.
Real-time classical-statistical simulations have been extensively used to study the dynamics of isolated systems in corresponding regimes of applicability for similar initial conditions \cite{Boguslavski:2019ecc, Noel2024:PhysRevD.109.056011, Mikheev:2024pur,Berges:2008wm, Gasenzer:2011by, Berges:2013eia, Berges:2013fga, Ewerz:2014tua, Maraga2015AgingCoarsening, Moore:2015adu, Berges:2016nru, Boguslavski:2019fsb, Nowak:2011sk, Schole:2012kt, Karl:2013kua, Hofmann2014a, Karl:2016wko, Williamson2016a.PhysRevLett.116.025301, Deng:2018xsk, Schmied:2018osf.PhysRevA.99.033611, Schmied:2019abm, Gresista:2021qqa, Heinen:2022rew, Siovitz:2023ius.PhysRevLett.131.183402, Spitz:2024jeg, Siovitz:2024aqi}.

% ==================================================================== 
\section{Excitations and defects in O(1) theory}
\label{sec:dof}
%
% ===========================================
\begin{figure*}[htb!]
    \centering
    \includegraphics[width=0.85\linewidth]{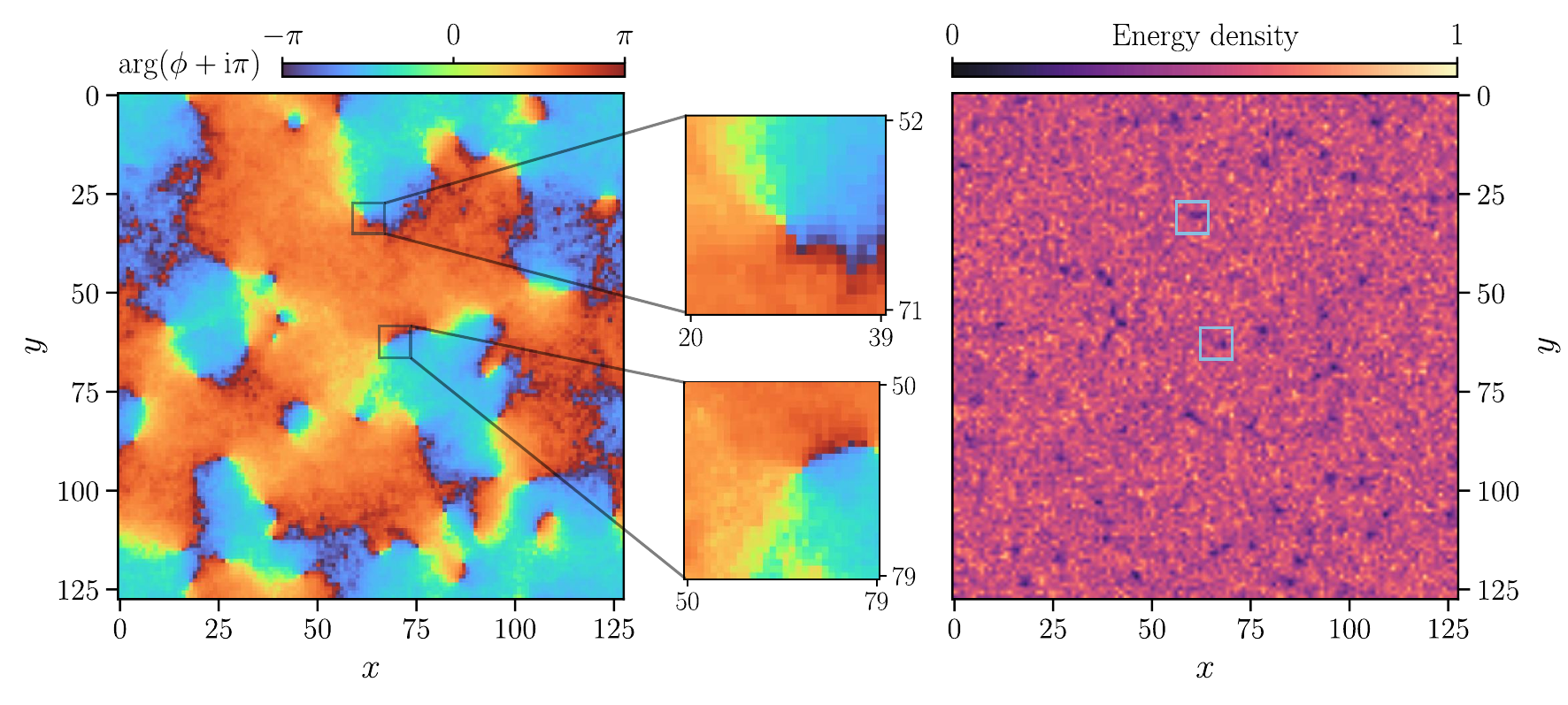}
 \caption{{\em Left:} Phase angle of the complex quantity $\theta=\phi+\mathrm{i}\pi$ defined from the relativistic degrees of freedom. Vortices/antivortices occur at a phase change $\pm 2 \pi$, as magnified by the middle insets. {\em Right:} Vortex defects also show up as distinct minima in the local energy density on the lattice. We show coarse-grained plots to minimise the influence of lattice artefacts. Lengths are shown in units of $Q^{-1}$, cf.~\Eq{boxinicond}. 
 The healing length $\xi_\mathrm{h}=p_{\xi_\text{h}}^{-1}=[2g\rho_0M]^{-1/2}$, which gives the scale of the diameter of the vortex cores, is $\xi_\text{h}\approx0.83\,Q^{-1}$.
 }
    \label{fig:defectsmp}
\end{figure*}
% ===========================================
In this section, we discuss the different quasiparticle and topological defect contributions that we expect to be relevant for $\mathrm{O}(1)$ scalar field theory. 

% ==================================================================== 
\subsection{Low-energy effective theories of scalar fields}
Relativistic $\mathrm{O}(N)$ scalar field theories are known to display nonrelativistic effects at low momenta in the presence of a mass gap. 
For the initial conditions considered here, such a mass gap $|p| \lesssim M$ with effective mass $M$ emerges even when the underlying theory is originally massless \cite{PineiroOrioli:2015cpb}, giving rise to nonrelativistic dynamics in the infrared.
Hence, a single component field theory $\mathrm{O}(1)$ can be described by an emergent $\mathrm{U}(1)$ theory. 
This mapping can be made more rigorous by introducing nonrelativistic degrees of freedom emerging from the underlying relativistic ones 
\begin{equation}
    \label{eq:nonrel_transform}
    \psi=\e^{\i M t}\left[\sqrt{\omega_x}\, \phi+\i / \sqrt{\omega_x}\, \pi\right] / \sqrt{2}\,,
\end{equation}
with 
$\omega_x=\sqrt{M^2-\nabla^2} $ \cite{namjoo2018relativistic, Deng:2018xsk}.
The emergent dynamics is governed by the Gross-Pitaevskii equation,
\begin{equation}
\i \partial_t \psi(t, \mathbf{x})=-\frac{\nabla^2}{2 M} \psi(t, \mathbf{x})+g|\psi(t, \mathbf{x})|^2 \psi(t, \mathbf{x})\,,
\label{eq:GPE}
\end{equation}
thus describing the real-time dynamics of a single-component Bose gas. 
While it is a well-known mapping, Ref.~\cite{Deng:2018xsk} considered it for the case of overoccupied isolated systems corroborated by numerical simulations in two dimensions, with initial conditions very similar to what is being considered here. 

In contrast, for a nonrelativistic $N$-component Bose gas, a low-energy effective theory was considered in Ref.~\cite{Mikheev:2018adp}, describing interacting Goldstone modes of the total and relative phase excitations. 
There are $N$ equations of motion for the $\mathrm{U}(N)$ theory, which give rise to $N-1$ Goldstone modes\footnote{Based on the underlying $U(N)$ theory, there are $2N-1$ broken generators, and $2N-2$ of them combine into the $N-1$ quadratic modes \cite{Mikheev:2018adp}.} with quadratic dispersion $\omega_{\mathrm{G}}(p)$ (relative phase excitations), and a Bogoliubov mode (total phase) with dispersion $\omega_{\mathrm{B}}(p)$,
\begin{equation}
\label{eq:disp_nonrel}
\begin{aligned}
& \omega_{\mathrm{G}}(p)=\frac{p^2}{2 m}\,, \quad c=1, \ldots, N-1\,, \\
&  \omega_{\mathrm{B}}(p)=\sqrt{\frac{p^2}{2 m}\left(\frac{p^2}{2 m}+2 g \rho_0\right)}\,,
\end{aligned}
\end{equation}
where $\rho_0$ is the condensate density in the zero-mode.
In Ref.~\cite{Boguslavski:2019ecc}, these dispersions were generalised to the relativistic case of O($N$) models by computing the dispersions of possible excitations using classical field equations of motion. This gives rise to the gapless modes 
\begin{align}
\label{eq:GRel}
\omega_{\mathrm{G,rel}}(p)&=\sqrt{p^2+M^2}-M \\
\label{eq:BogRel}
\omega_{\mathrm{B,rel}}(p)
&=\sqrt{p^2+2 M\left(M+g \rho_0 - \sqrt{p^2+\left(M+g \rho_0\right)^2}\right)}\,,
\end{align}
where $g \rho_0$ imprints the magnitude of the (local) condensate and is related to the effective mass $M$.
These dispersion relations are provided as relative excitations on top of the local condensate that is the solution of the classical equation of motion. Therefore, they can be directly compared to the dispersions \eqref{eq:disp_nonrel} of the nonrelativistic theories. However, we emphasise that in the relativistic case, one computes correlations of the scalar field $\phi$ instead of $\psi$ and, before comparing to the stated dispersions, one has to remove the rotation frequency $\pm M$ in the frequency spectrum \cite{Boguslavski:2019ecc} that also emerges in the exponent of Eq.~\eqref{eq:nonrel_transform}. We will take this into account in all the figures below by plotting $\omega - M$.

\begin{figure*}
    \centering
    \includegraphics[width=0.95\linewidth]{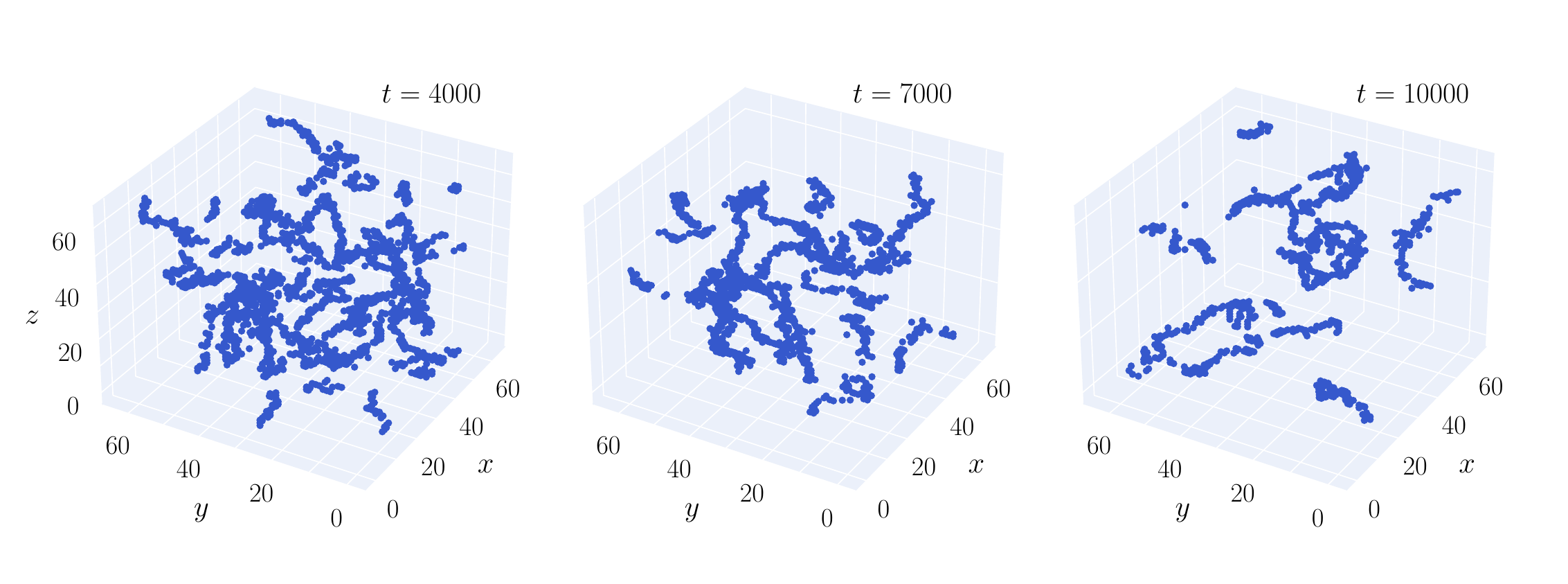}
    \caption{Tangles of vortex lines on a three-dimensional lattice shown at three different times and extracted from the low-energy depressions of the energy density fluctuations. Due to reconnections of vortex rings, leading to both larger and smaller rings and subsequent shrinking and dissolution of the smallest rings on the noisy background, the overall vortex line length decreases, corresponding to a coarsening process in the energy density.
    Lengths are shown in units of $Q^{-1}$.
    The healing length is $\xi_\mathrm{h}=p_{\xi_\text{h}}^{-1}=[2g\rho_0M]^{-1/2}\approx1.67\,Q^{-1}$.
    }
    \label{fig:defectsmp2}
\end{figure*}

We note that in the large-$N$ limit, the Bogoliubov mode is suppressed in the infrared. 
In this limit, large-$N$ kinetic theories \cite{PineiroOrioli:2015cpb, Walz:2017ffj, Chantesana:2018qsb} that dominantly describe the scattering of quasiparticles with a quadratic dispersion agree with the low-energy effective theory (EFT) considered in Ref.~\cite{Mikheev:2018adp}. For the single-component scalar field theory, the EFT considers only Bogoliubov modes and no quadratic excitations. 
However, the extraction of the excitation spectrum in $\mathrm{U}(1)$ and $\mathrm{O}(1)$ theories in Refs.~\cite{PineiroOrioli:2018hst, Boguslavski:2019ecc} reveals the existence of a dominant ``transport peak'' or ``non-Lorentzian peak'' at low momenta, in addition to the Bogoliubov modes, that drives the dynamics.
This raises the question of what kind of quasiparticles are present in a single-component scalar field theory and, in particular, what the dominant degree of freedom is that none of the existing EFTs or kinetic theories are able to capture. 
% ==================================================================== 
\subsection{Topological defects in $\mathrm{O}(1)$ scalar field theory}
Approaching from a different perspective, an important consideration is the potential existence of topological defects in our systems.
This could be especially significant for the $N=1$ case, since the lower $N$ is, the more constrained the topology of the underlying vacuum manifold becomes, making it more likely to support nontrivial topological structures.%
\footnote{For a mathematically more rigorous consideration of what plays the role of a vacuum manifold in the nonequilibrium dynamical theory considered here, see Refs.~\cite{Moore:2015adu, Noel2024:PhysRevD.109.056011}. }

Due to the effective conservation of particle number in the infrared regime, which also occurs in the presence of a mass gap in the relativistic theories considered here, an inverse self-similar particle transport develops. 
This gradually transfers particles to low-momentum modes, ultimately building a condensate at zero momentum. 
However, causality imposes a limit on the rate at which information and correlations can propagate, preventing the instantaneous formation of a uniform condensate.
Therefore, the condensate initially lacks long-range order, which it will gradually develop through ordering dynamics \cite{Bray1994a.AdvPhys.43.357}. 
This dynamics may include topological defects depending on the number of field components $N$ \cite{Moore:2015adu}, and for a single component, $N=1$, there are stable vortex defects. 
In this particular case, the formation of vortices and antivortices, and their annihilation process is of importance, since this allows for the growth of larger structures.

One way to identify these defects in our simulations would be performing the canonical transformation \eqref{eq:nonrel_transform} on the relativistic degrees of freedom $\phi$ and $\pi$ to obtain the complex $\psi(t, \mathbf{x})$. 
Then positions on the real-space lattice where depressions in the density of $|\psi(t, \mathbf{x})|^2$ occur along with a phase change of approximately $\pm 2 \pi$ in $\arg [\psi(t, \mathbf{x})]$ would correspond to vortices and antivortices, respectively \cite{Deng:2018xsk}.
However, we find that it is also possible to obtain topological defects directly from the relativistic degrees of freedom $\phi$ and $\pi$ without transforming into $\psi$, and also without applying any cooling effects in the simulations.
For a relativistic theory with a sufficiently large mass gap, defects also show up as minima in energy density fluctuations \cite{Noel2024:PhysRevD.109.056011}, which coincide with a $\pm 2 \pi$ winding in the relativistic phase angle defined by 
\begin{equation}
    \theta(t,\mathbf{x}) = \mathrm{arg}\left[\phi(t,\mathbf{x}) + \i\, \pi(t,\mathbf{x})\right]\,,
\end{equation} as shown in Fig.~\ref{fig:defectsmp} for the two-dimensional case.
We have cross-checked that the number and positions of (anti)vortices coincide in these two observables, the relativistic $\theta(t,\mathbf{x})$, and the nonrelativistic  $\arg [\psi(t, \mathbf{x})]$.
Such phase singularities in the nonrelativistic $\mathrm{arg}[\psi(t, \mathbf{x})]$ have been specifically matched with depressions in the complex field $\psi$ in Refs.~\cite{Karl:2016wko, Deng:2018xsk} and their time evolution has been tracked in the context of vortex-antivortex annihilation events and the resulting coarsening of the domains. 
In this work, we will obtain results from purely relativistic quantities, namely the fields $\phi(t,\mathbf{x})$ and $\pi(t,\mathbf{x})$ entering $\theta(t,\mathbf{x})$. 

Technically, one could first identify the minima in the energy density and check for a corresponding phase winding in $\theta(t,\mathbf{x})$. In practice, however, nearly all the $\pm 2\pi$ phase windings coincide with a minimum in the energy density, such that simply checking only one of these observables gives an accurate picture of the total vortex density.
Similarly, for the three-dimensional case, vortex lines are revealed by looking for depressions in the energy density fluctuations, as seen in Fig.~\ref{fig:defectsmp2} for different times.

% ===========================================
\begin{figure}[t]
    \centering
        \includegraphics[width=\linewidth]{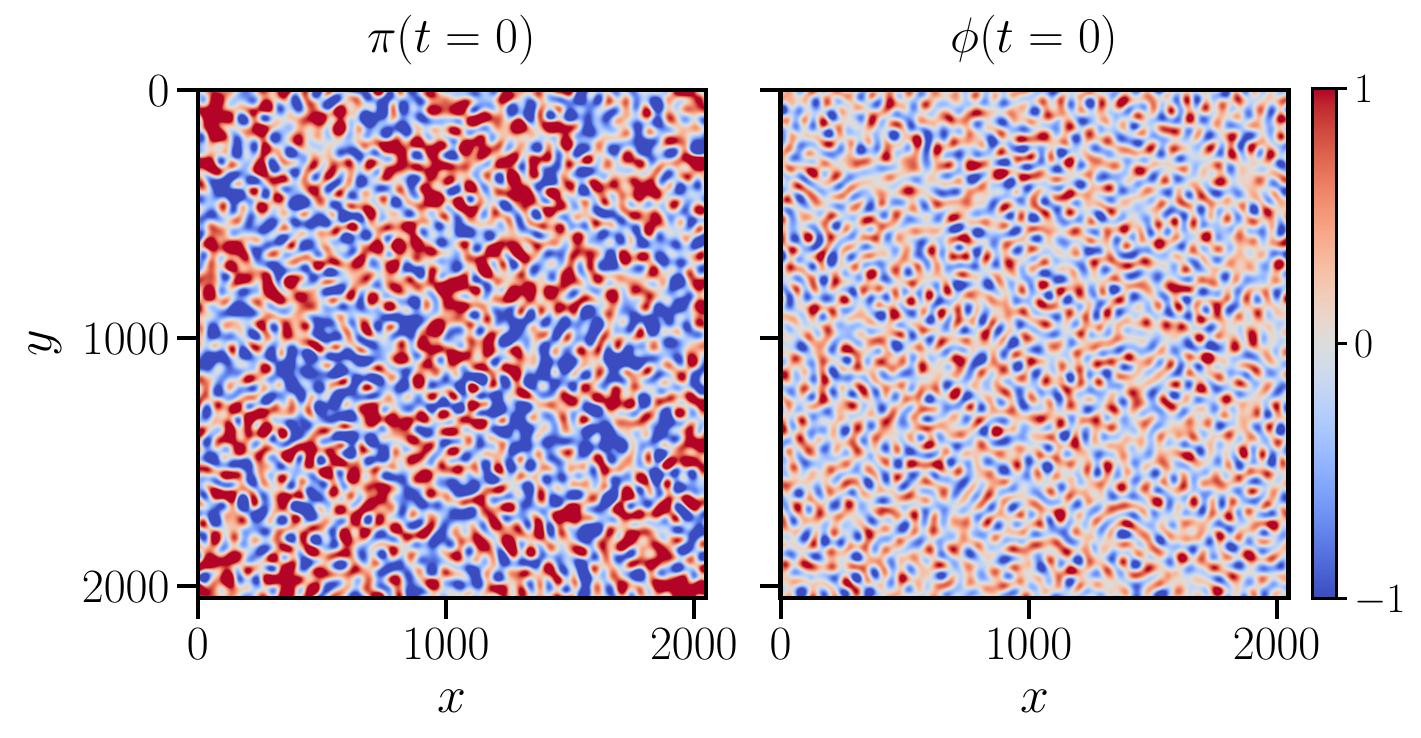}
        \caption{Initial spatial configurations of the fields $\phi(t_0,\mathbf{r})$ and $\pi(t_0,\mathbf{r})$ in $d=2$ dimensions, $\mathbf{r}=(x,y)$, normalised to have values between $-1$ and $1$. 
        The ``scrambled'' pattern, containing many small domains, allows for numerous phase singularities in $\arg [\phi+\i \pi]$, which we identify as (anti)vortex defects. These domains then become larger, as throughout the coarsening process vortices annihilate with antivortices resulting in the growth of structures.
        The healing length is $\xi_\text{h}\approx0.83\,Q^{-1}$ in the units shown.}
        \label{fig:phi_pi}
\end{figure}
\begin{figure*}
    \centering        \includegraphics[width=0.96\linewidth]{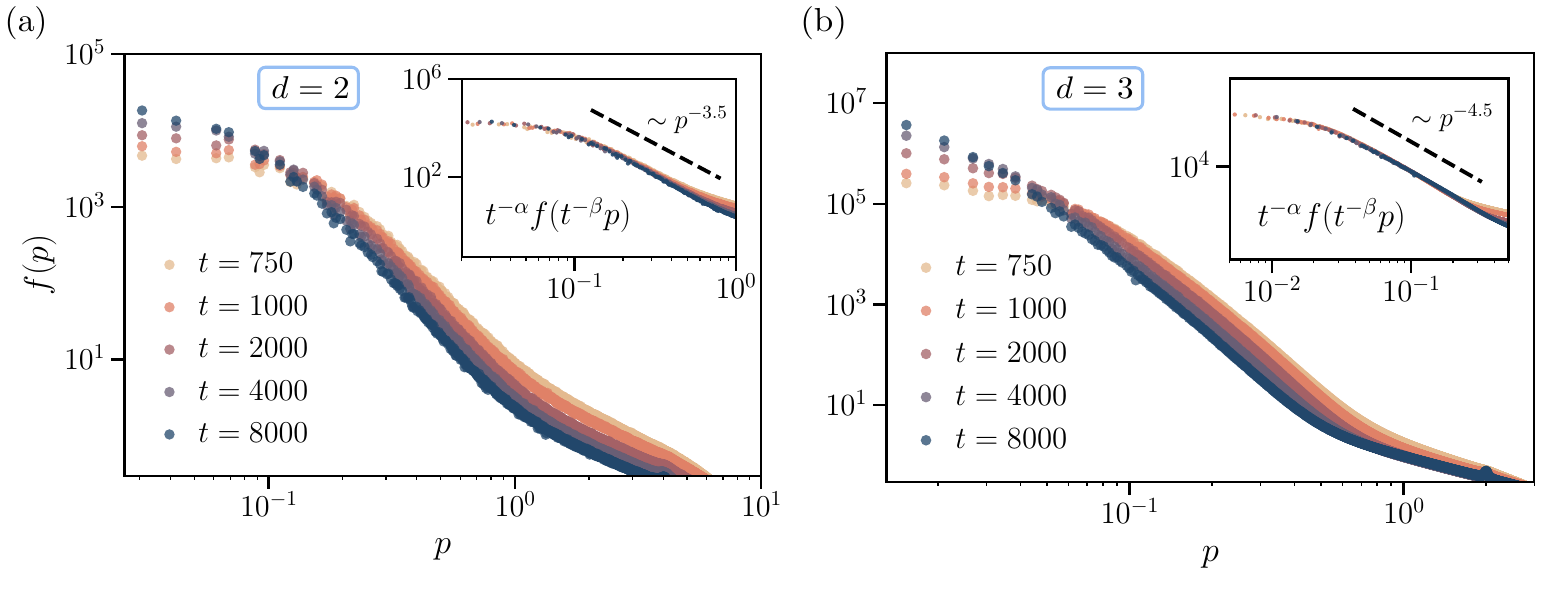}
        \caption{Distribution function $f(t,p)$, cf.~\eq{Defftp}, for a single-component scalar field theory in (a) two and (b) three dimensions, shown at five different evolution times $t$, where quantities are shown in units of $Q$, i.e., time in units of $Q^{-1}$ and momentum in units of $Q$. The inset shows the functions rescaled according to \eq{fselfs}, with $\beta=0.25$ and $\alpha=0.5$ in two dimensions, and $\beta=0.5$ and $\alpha=1.5$ in three dimensions. We also show approximate power law fits $p^{-\kappa}$ in both cases. In (a), simulations were run on a $2048^2$ lattice, while in (b) the lattice size was $512^3$.
        The momentum scales $p_{\xi_\text{h}}=\sqrt{2g\rho_0M}$ corresponding to the healing length $\xi_\mathrm{h}=p_{\xi_\text{h}}^{-1}$ are $p_{\xi_\text{h}}\approx1.2\,Q$ in $d=2$ and $p_{\xi_\text{h}}\approx0.6\,Q$ in $d=3$.
        }
        \label{fig:distribution}
\end{figure*}
% ===========================================

Since vortex configurations are solutions to the classical equation of motion, they can arise during the dynamics. 
However, it is worth mentioning that even though these vortices are not explicitly part of the initial conditions under consideration, our box initial condition \eq{boxinicond} provides the right environment for vortices to emerge.
This can be seen in Fig.~\ref{fig:phi_pi}, where the initial field configurations $\phi(t=0,\mathbf{x})$ and $\pi(t=0,\mathbf{x})$ are shown in position space. 
The ``scrambled'' look results from small domains that are characterised by sharp boundaries between low and high values of $\phi$ and $\pi$ due to the step function in the distribution at the characteristic momentum $Q$. 
As already illustrated in Fig.~\ref{fig:defectsmp}, such sharp boundaries are ideal for phase singularities in $\theta(t=0,\mathbf{x})$, which is a feature of vortices. 

It is also important to note that using a smoother cutoff for the initial momentum space distribution instead of a step function, like a hyperbolic tangent function or a Gaussian function, did not reduce the number of vortex defects initially formed in the system, which we explicitly checked.
As long as one has a relatively steep increase in occupation number in momentum space near some characteristic momentum, this initially scrambled picture will emerge in position space, and vortices will always form very quickly in the dynamics.
In this sense, it is not possible not to have, or to remove these defect configurations.
For the case of $N=1$ (and also $N=2,3$ but not higher $N$), they are stable configurations on topological grounds \cite{Moore:2015adu, Noel2024:PhysRevD.109.056011}.

% ==================================================================== 
\subsection{Excitations on vortices}
Another type of contribution yet to be considered is that the vortex defects themselves can support excitations, which appear as quasiparticles during the dynamics.
For instance, phonons can be produced from vortex-antivortex annihilation processes \cite{barenghi2001quantized, richaud2023mass}, which are expected to yield a linear dispersion relation at low momenta. 
Beyond this, individual vortex lines in three dimensions, and vortex cores in two dimensions can be excited in a way that can be characterised as quasiparticles, which are distinct from the phonons arising from vortex annihilation.
In anticipation of our findings, we briefly summarise the properties of such \textit{kelvon} quasiparticles.

In three dimensional systems, vortex defects are extended objects and may be viewed as vortex lines, along which helical excitations can propagate as waves.
These are known as Kelvin waves \cite{Thomson1880}, and were originally described in the context of classical hydrodynamics.
In the realm of quantum fluids, Kelvin waves arise as helical perturbations of quantised vortex lines, with the circulation being discretised as a direct consequence of the quantum mechanical phase winding around the vortex core \cite{Pitaevskii1961a,RobertsPH2003a.rspa.459.331,RobertsPH2003b.rspa.459.597,Donnelly2005a}.
These excitations play a crucial role in the dynamics of quantum turbulence, mediating energy transfer from larger to smaller scales \cite{vinen2001decay,vinen2003kelvin,kozik2004kelvin,Kozik2005a.PhysRevLett.94.025301,kozik2005vortex,Kozik2008a.PhysRevLett.100.195302,Kozik2009a, Lvov2010a, Krstulovic2012a.PhysRevE.86.055301,Laurie2010a.PhysRevB.81.104526, Boue2011a.PhysRevB.84.064516, boue2015energy}, as Kelvin waves at a similar scale interact, exciting smaller scale Kelvin waves, ultimately dissipated by phonon emission \cite{vinen2003kelvin, kozik2005vortex, vinen2001decay}.
At larger scales, vortex reconnection is thought to dominate energy transfer \cite{meichle2012quantized}, and such reconnection events excite Kelvin waves at scales close to the inter-vortex distance. 
Lord Kelvin obtained, for the small-amplitude wave excitations of a thin columnar vortex with a hollow core, with wave number $p$ along the vortex line and azimuthal quantum number $n=1$, the dispersions \cite{Thomson1880}
\begin{equation}
\omega^{\pm}_\mathrm{K}(p)
=\frac{\Gamma}{2\pi r_\mathrm{c}^2}
\left(1 \pm \sqrt{1+r_\mathrm{c} p  \frac{K_0(r_\mathrm{c} p )}
{K_1(r_\mathrm{c} p)}}\right)\,,
\label{eq:kelvon}
\end{equation}
where $r_\mathrm{c}$ is the vortex core radius, $\Gamma=\oint_{\,\mathcal{C}}\mathbf{v}\cdot\mathrm{d}\mathbf{r}$ the vortex circulation, with fluid velocity $\mathbf{v}$ and $\mathcal{C}$ some closed path around the core, and $K_j$ are modified Bessel functions of order $j$.
The low-momentum approximation of the dispersion $\omega^-$ has the form
\begin{equation}
\omega^-_\mathrm{K}(p)
\simeq-\frac{\Gamma p^2}{4\pi}\left[\ln\left(\frac{2}{pr_\mathrm{c}}\right)-\gamma\right]\,,
\quad (pr_\mathrm{c}\ll1)\,,
\label{eq:kelvonlongwavelength}
\end{equation}
where $\gamma\approx0.577216\dots$ is the Euler-Mascheroni constant.
The Kelvin-wave dispersion \eq{kelvonlongwavelength} also applies to vortices in a superfluid \cite{Donnelly2005a} with integer winding number ${w}\in\mathbb{Z}$ and circulation $\Gamma={w}h/m$ quantised in integer multiples of the Planck constant $h$ over particle mass. 
The much larger $\omega^+(p\to0)\simeq\Gamma/(\pi r_\mathrm{c}^2)$ is related to the Magnus force and the mass inside the core \cite{Giuriato2020a.PhysRevResearch.2.023149, richaud2021dynamics}.

The dynamics of quantised vortices and their Kelvin-wave excitations with dispersion \eq{kelvon} is well captured by the Gross-Pitaevskii equation (GPE) \eq{GPE} \cite{simula2008kelvin, simula2008vortex,simula2013collective, baggaley2014kelvin, clark2015spatiotemporal}.
As it is explicitly visible in the approximate expression \eq{kelvonlongwavelength}, the $p$-dependence of the dispersion \eq{kelvon} is nearly quadratic at low momenta. 
Around $pr_\mathrm{c}=1$ it is close to linear in $p$, while at higher momenta, it bends to a lower power than linear.
Such a momentum dependence serves as a unique signature of Kelvin waves that is unlikely to be attributed to any other excitation with only linear or quadratic dispersions.
Previous GPE studies of Kelvin waves mostly focused on weak wave turbulence, and a direct connection has not been made with nonthermal fixed points to date, where strong turbulence, including vortex tangles, dominates the infrared region.

At first sight, Kelvin waves cannot occur in two dimensions, where vortices lack a linear extension along the core and are localised objects with circular symmetry. 
Similar quasiparticles, however, emerge even in two dimensions  \cite{simula2018vortex, simula2020gravitational}, especially if the vortices are not point-like, but have a finite core size, which is precisely our case, as seen in the right panel of Fig.~\ref{fig:defectsmp} depicting the energy density.
As a result, the \textit{boundary} of the vortex supports bound-state-like oscillations, which, in the azimuthal direction, are quantised in the same way as kelvons in three dimensions, looked at in a single fixed plane perpendicular to the vortex core.
The lowest such azimuthal excitation corresponds to a circularly varying displacement of the centre of the vortex.
As a result, also the density is modified in its radial dependence relative to the rotating position of the vortex, and in the large-volume limit, a continuum of kelvon states emerges, which can be distinguished again by a wave number $p$ parametrising the radial dependence as well as the dispersion of the kelvon.
Therefore, an analogy can be drawn between a Kelvin wave extended \textit{along} a vortex line in three dimensions and a ``Kelvin wave'' representing, at lowest order, a circularly rotating displacement of the vortex position.
We will also refer to these quasiparticles in two dimensions as kelvons, for which a dispersion with the same scaling \eq{kelvonlongwavelength} as in three dimensions has been found \cite{simula2018vortex, simula2020gravitational}.
Wave-like motion along vortex boundaries has also been discussed in the context of Kelvin waves in finite-size, quasi two-dimensional systems as realised in a trapping potential strongly confined in the third dimension \cite{simula2018vortex, simula2013collective}.

% ==================================================================== 
% ==================================================================== 
\section{Results: Universal scaling dynamics}
\label{sec:results}

% ==================================================================== 
\subsection{Self-similar evolution of the momentum spectrum}
\label{sec:EqualTimeCorrFunc}

The O$(1)$ theory \eq{OoneModel} in $d=2$ spatial dimensions provides a particularly intriguing and valuable framework for the study of relevant degrees of freedom for our initial conditions.  
Comparable simulations in $d=3$ for the massive O$(1)$ theory \cite{Boguslavski:2019ecc, Moore:2015adu} have revealed the momentum distribution function \eq{Defftp} which evolves self-similarly according to
\begin{equation}
f(t, p)=t^{\,\alpha} f_\mathrm{s}\left(t^{\,\beta} p\right)\,,
\label{eq:fselfs}
\end{equation}
with universal scaling exponents $\alpha$ and $\beta$ and scaling function $f_\mathrm{s}(p)$. 
The distribution function takes the approximate form
\begin{equation}
    f(t, p) \approx \left(\frac{\mathcal{N}(t)}{\pL^2(t)+p^2}\right)^{\kappa/2}\,,
\end{equation}
where the characteristic momentum scale $\pL$ marks the transition between a plateau in the infrared and power law fall-off $p^{-\kappa}$.
For $p\to0$, the function develops a plateau or increases with a power law much smaller than $\kappa$.

According to \Eq{fselfs}, the characteristic scale decreases in time as $\pL(t)\sim t^{-\beta}$, while the overall normalisation scales as $\mathcal{N}(t)\sim t^{2(\alpha/\kappa-\beta)}$.
If particle number and thus the integral $\int\text{d}^dp f(t,p)$ is conserved in time, in $d$ dimensions, the relation $\alpha=d\beta$ must hold.
Semiclassical simulations gave the universal scaling exponents  $\beta \approx 0.5$ and $\alpha \approx 1.5\approx d\,\beta$, and $\kappa\approx 4.5$ in $d=3$ dimensions \cite{PineiroOrioli:2015cpb, Moore:2015adu}. 

This self-similar scaling is confirmed by our results in Fig.~\ref{fig:distribution}b, where we show the evolution of $f(t,p)$ in $d=3$ dimensions for different times as a function of momentum. 
One finds that the characteristic self-similar dynamics in Eq.~\eqref{eq:fselfs} that builds up a condensate, i.e., a macroscopic occupation of the low-$p$ modes, requires the stated values for $\alpha$ and $\beta$ to obtain a time-insensitive scaling function $f_\mathrm{s}$ as visible in the inset. 
This form consists of a nearly constant part at the lowest momenta $p \lesssim \pL$ and the power law decrease $p^{-\kappa}$ at higher momenta $p \gtrsim \pL$ within the infrared momentum range below the inverse healing length $p \lesssim p_{\xi_\text{h}} = \sqrt{2g\rho_0M}$.

% ===========================================
\begin{figure}[t]
    \centering
    \includegraphics[width=0.92\linewidth]{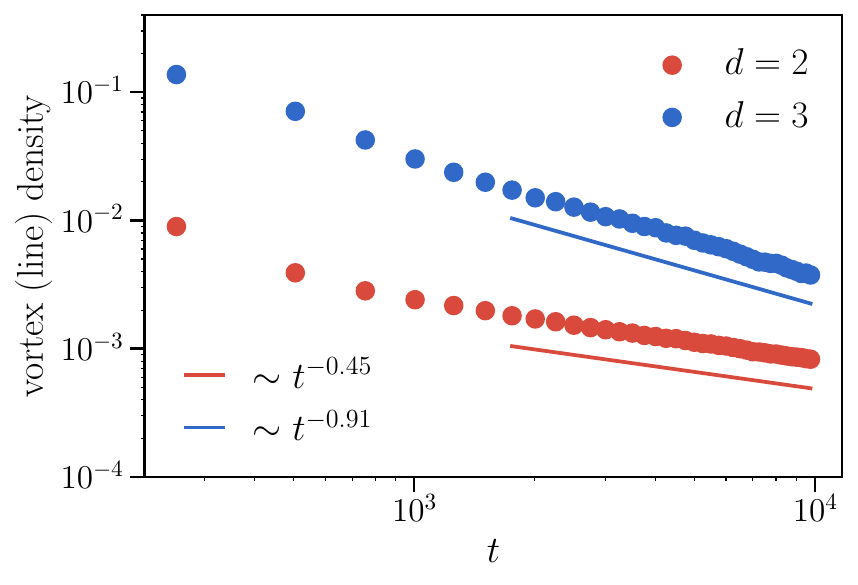}
    \caption{Vortex (line) density shown as a function of time, in $d=2$ and $d=3$ dimensions for O$(1)$ scalar fields. The decay reveals a power law $N_\mathrm{v}\sim t^{-\xi}$ with $\xi\approx0.45$ in two dimensions, and $\xi\approx0.91$ in three dimensions.}
    \label{fig:vortex}
\end{figure}
% ===========================================

% ==================================================================== 
\subsection{Coarsening dynamics}
\label{sec:Coarsening}
Based on the spectral and statistical functions, the above excitation dynamics for single-component scalar fields was found to be markedly different from what is observed and expected in the large-$N$ limit \cite{Boguslavski:2019ecc}, where we have an analytic understanding of these exponents \cite{PineiroOrioli:2015cpb, Walz:2017ffj, Chantesana:2018qsb}. 
On the other hand, the exponent $\beta \approx 0.5$ is known to apply in qualitatively very different situations. 
As an alternative to the self-similar transport of free Goldstone quasiparticle excitations to lower momenta, which dominates the large-$N$ limit \cite{Chantesana:2018qsb, Mikheev:2018adp}, it could also result from diffusion-type coarsening dynamics of the spatial field configuration, e.g., as induced by mutual annihilation of vortices and antivortices excited in the system \cite{Karl:2016wko}.
 
Yet, in two-dimensional O$(1)$ simulations, the equal-time distribution function has only been found to scale with $\beta \approx 0.25$, $\alpha \approx 0.5$, and $\kappa\approx4$ \cite{Deng:2018xsk}. 
The temporal scaling exponents $\alpha$ and $\beta$ are confirmed by our results in Fig.~\ref{fig:distribution}a, where we show the evolution of $f(t,p)$ for a two-dimensional system in the main panel and its self-similar scaling in the inset using the stated exponents. We find the power law decrease of the distribution function to be closer to $\sim p^{-3.5}$ for the relatively short momentum interval shown, and we do not exclude the previously observed value of $\kappa$.
Despite scanning through a wide range of initial conditions, these results remained unchanged. In particular, we were unable to find a stable $\beta \approx 0.5$ scaling regime in $d=2$. In contrast, for $N>1$, we have easily recovered the diffusion-type scaling with $\beta \approx 0.5$ and $\alpha = d \beta \approx 1.0$.

% ===========================================
\begin{figure*}[t]
    \centering
    \includegraphics[width=0.95\linewidth]{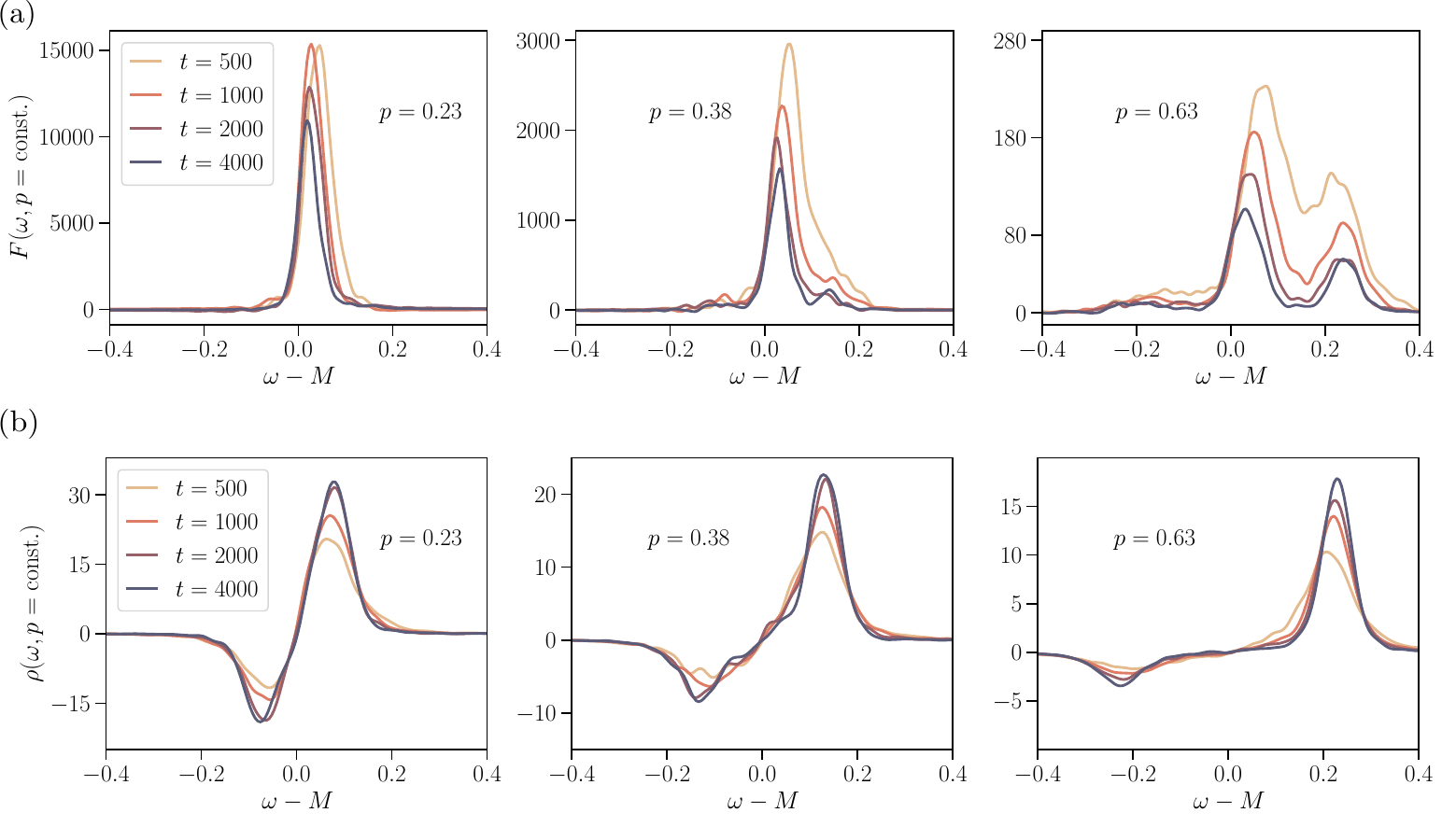}
    \caption{Fourier transform, \eq{FTDefFtauomegap}, of the (a) unequal-time statistical function, $F(t,\omega,p)$, and the (b) spectral function $\rho(t,\omega, p)$ at different central times $t$ and momenta $p$ in $d=2$ dimensions. They show the dominant transport peak in the vicinity of $\omega-M \approx 0$ in $F(t,\omega,p)$, with both branches of the Bogoliubov quasiparticle peak emerging to the left and right as larger momenta are considered. The spectral function is dominated by the Bogoliubov peak at all momenta. The curves are shown as functions of $\omega-M$, with zero-momentum mass gap $M$, and momenta and frequency in units of $Q$.
    }
    \label{fig:Ffun}
\end{figure*}
% ===========================================
A closely related scaling exponent to the $d=2$ case of $\mathrm{O}(1)$ theory, $\beta\approx 0.2$ has been found in simulations of a nonrelativistic Bose gas in two dimensions upon imprinting vortices into the initial field configuration \cite{Karl:2016wko}, also confirmed experimentally \cite{Johnstone2019a.Science.364.1267}.
This scaling behaviour is phenomenologically understood as due to  three-body collisions of bound vortex-antivortex pairs with nearby vortices or antivortices.
Any such collision allows for an exchange of vortices after which the newly formed vortex-antivortex pair can be more strongly bound and thus potentially undergo fast subsequent vortex-antivortex annihilation \cite{Karl:2016wko}.
Hence, pair annihilation is dominated by the three-body collision rate, which gives rise to the strongly subdiffusive exponent $\beta\approx0.2\ll0.5$.
This stands in contrast to isolated vortex-antivortex annihilation due to background fluctuations, which is governed by a diffusion-type scaling with $\beta \approx 0.5$ \cite{Karl:2016wko,Baggaley2018a.PhysRevA.97.033601,Johnstone2019a.Science.364.1267,Groszek2021a.PhysRevResearch.3.013212}.
The relevance of three-body collisions was likewise identified in Ref.~\cite{Deng:2018xsk} for the $\mathrm{O}(1)$ model in two dimensions without imprinting vortices. 

In both Refs.~\cite{Karl:2016wko, Deng:2018xsk} the average vortex density was found to decay as a power law, $n_\mathrm{v}\sim t^{-\xi}$ with $\xi\approx0.41\approx2\beta$, which reflects that the decreasing infrared scale $\pL(t)\sim t^{-\beta}$ corresponds to a growing characteristic length scale, which can be associated with the mean inter-defect distance $\ell_\mathrm{v} \sim \pL^{-1}\sim t^{\beta}$.
Hence, it is useful to define the exponent 
\begin{equation}
    \xi=2\,\beta_{\mathrm{L}}\,,
\end{equation}
which thus characterises the growth of the average free volume between vortex defects in $d=2$ dimensions. Assuming that the self-similar scaling of the distribution function, \Eq{fselfs}, is related to the coarsening of topological defects, one has that $\beta_{\mathrm{L}}=\beta$.

An interesting question therefore is to what extent the spatiotemporal scaling of $f(t, p)$ and the momentum power law fall-off is, in general, driven by coarsening, or whether the two scaling phenomena are independent.
After all, the value of the exponent $\kappa$ and thus the steepness of the distribution determines whether and how particles can be transported towards the infrared and eventually allow condensation in the $p=0$ mode. 
The buildup of such a transport process necessarily involves some underlying coarsening due to causality, as explained in \Sect{model}.
As indicated by the numerical simulations in \cite{Mikheev:2018adp}, different processes driven by either, e.g., topological defects, or compressible sound excitations, can dominate the overall scaling evolution.

Hence, before moving on to investigating the excitation spectrum from unequal-time correlation functions, we examine the dynamics of vortices in two dimensions and vortex lines in three dimensions explicitly and extract the respective $\beta_{\mathrm{L}}$ exponents from the spatial field configurations. 
As mentioned in \Sect{dof}, (anti)vortex defects are identified by a $\pm 2\pi$ phase winding in $\theta(t,\mathbf{x})$, which coincide with a depression in the energy density on the lattice. 
Based on this, we have extracted the total vortex (line) density in two and three dimensions, which we show as functions of time in Fig.~\ref{fig:vortex}.
For additional details on the data analysis and the extraction of these quantities, see \App{vortdet}.

This vortex density reveals a power law $\sim t^{-\xi}$ with $\xi\approx 0.45$ or $\beta_{\mathrm{L}}\approx0.23$ for the $\mathrm{O}(1)$ model in two dimensions, which is consistent with the $\beta\approx 0.25$ self-similar scaling of the equal-time distribution function $f$.%
\footnote{Note however that, in two dimensions, vortex coarsening could also be characterised by $\beta_\mathrm{L}\approx\beta\approx 0.5$, as seen for a nonrelativistic Bose gas in \cite{Karl:2016wko} and discussed above.}
In three dimensions, we find $\xi\approx0.91$.
It represents the decrease of the total vortex line length per unit volume, which corresponds to a dilution of the vortex rings and tangles as $\sim\ell_\text{v}^{-2}\sim t^{-2\beta}$ \cite{Vinen1957a,Villois2016a.PhysRevE.93.061103},
where $\ell_\text{v}$ is the scale measuring the mean distance between vortex lines.
Hence, also in $d=3$, one has $\xi=2\beta$, and the decrease of the vortex line density implies $\beta_{\mathrm{L}}\approx0.46\approx\beta$.
While these numbers look convincing, a clearer link to the relevant quasiparticle excitations in $f$ would be desirable. 
To make such a connection, in the next subsection we discuss the extraction of the frequency and momentum dependence of the correlation functions that are closely related to the distribution function \eqref{eq:Defftp} but additionally encode the systems' excitation spectra.

% ===========================================
\begin{figure*}[t]
    \centering   \includegraphics[width=0.95\textwidth]{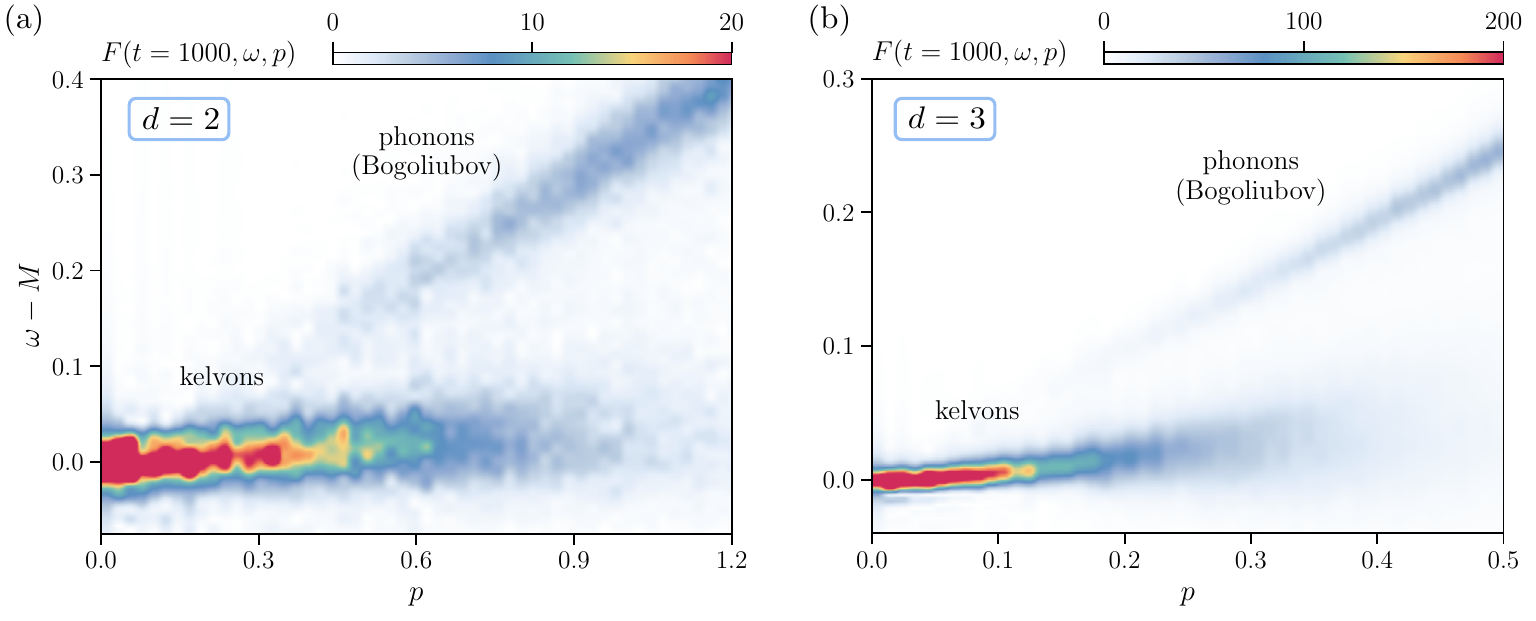}
    \caption{Frequency-momentum resolved representation of the statistical two-point correlator $F(t=1000,\omega,p)$ in (a) two and (b) three dimensions, normalised by the occupation number $f(t,p)$. 
    Momenta and frequencies are shown in units of $Q$.
    Note that the momentum scales $p_{\xi_\text{h}}=\sqrt{2g\rho_0M}$ corresponding to the healing length $\xi_\mathrm{h}=p_{\xi_\text{h}}^{-1}$ are $p_{\xi_\text{h}}\approx1.2\,Q$ in $d=2$ and $p_{\xi_\text{h}}\approx0.6\,Q$ in $d=3$.
    }
    \label{fig:omega_p_resolved}
\end{figure*}
% ===========================================
% ===========================================
\begin{figure}
    \centering
    \includegraphics[width=0.95\linewidth]{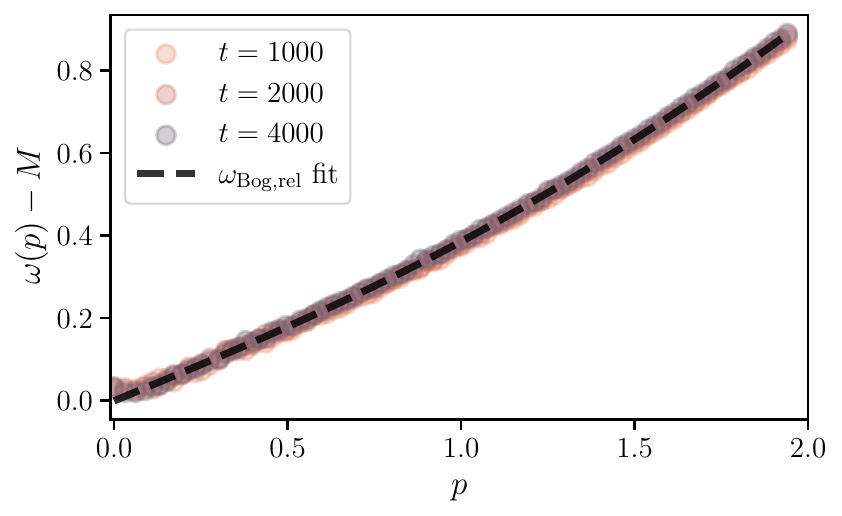}
    \caption{The dispersion relation extracted from the Bogoliubov peak in the two-dimensional spectral function. The fit corresponds to the relativistic generalisation of the Bogoliubov dispersion \eqref{eq:BogRel} and follows a linear, then quadratic, and finally again linear evolution.
    The healing-length momentum scale is $p_{\xi_\text{h}}\approx1.2\,Q$.
    }
    \label{fig:bogdisp}
\end{figure}
% ===========================================
% ===========================================
\begin{figure*}[t]
    \centering
    \includegraphics[width=0.94\textwidth]{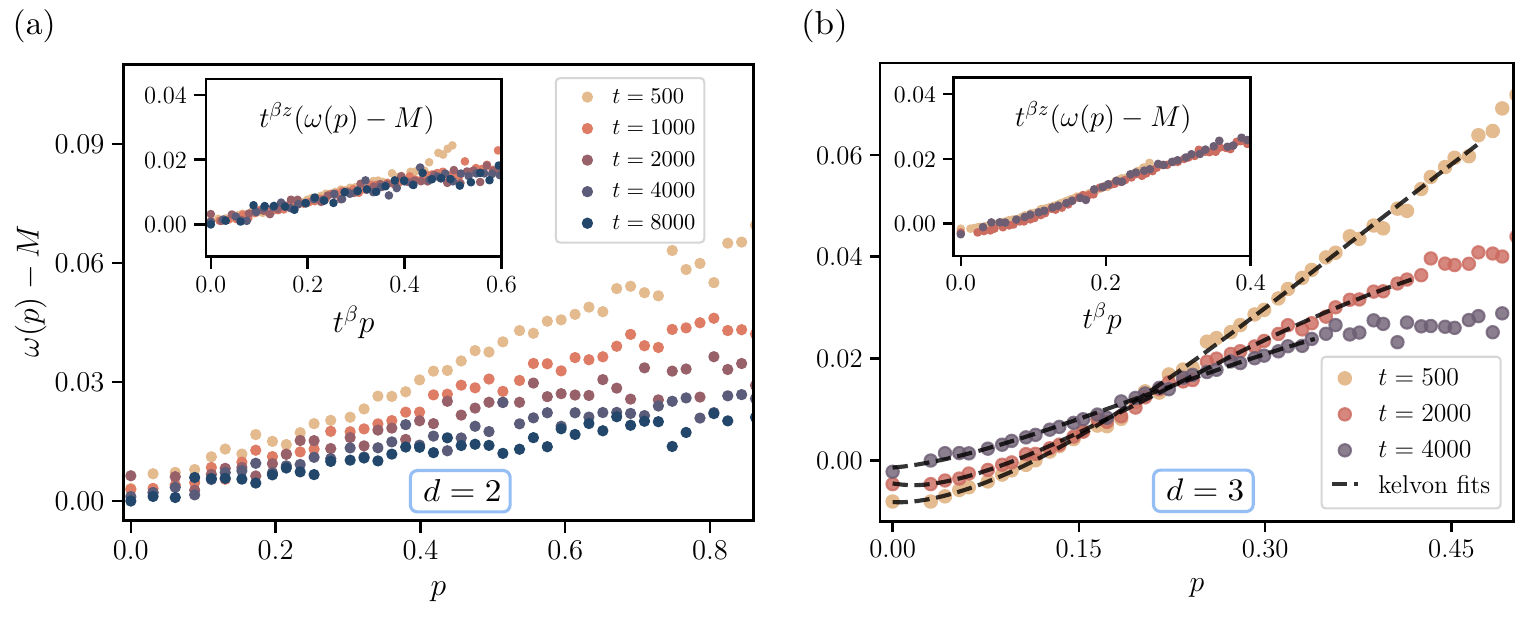}
    \caption{Time-dependent dispersion relations $\omega_\mathrm{K}(p)$ of the infrared transport peak for a single component scalar field theory: (a) In two dimensions, $\omega_\mathrm{K}(p)$ approximately follow straight lines. (b) In three dimensions, the curves are accurately fitted by the Kelvin wave dispersion relation \eq{kelvon2}. Here we show only 3 curves to make it visually clear how the data is captured by the fit. The insets show the rescaled dispersions. In two dimensions, we used $\beta=0.25$, $z=2$, while in three dimensions $\beta=0.5$, $z=2$. 
    }
    \label{fig:combined_dispersion}
\end{figure*}
% ===========================================
% ==================================================================== 
\subsection{Unequal-time correlation functions}
A slow self-similar scaling dynamics of the equal-time distribution function $f(t,p)$ with $\beta\approx0.25$ for the $\mathrm{O}(1)$ theory has only been found in two dimensions. 
Therefore, this provides an interesting study case to investigate the unequal-time correlation functions in both two and three dimensions, focusing on understanding the differences in scaling exponents and also on the potential connections to the dynamics arising from topological defects. 

For both the two- and three-dimensional theories, the statistical function $F$ is dominated by a single peak at low momenta in the infrared, which is accurately captured by the (non-Lorentzian) hyperbolic secant function,%
\footnote{Regarding the notation: this peak corresponds to the ``transport peak'' in \cite{PineiroOrioli:2018hst} and the ``non-Lorentzian peak'' (nL) in \cite{Boguslavski:2019ecc}.}
\begin{equation}
F_{\mathrm{K}}(t, \omega, p) \simeq \frac{\pi}{2} \frac{A_{\mathrm{K}}(t, p)}{\gamma_{\mathrm{K}}(p)} \operatorname{sech}\left[\frac{\pi}{2} \frac{\omega-\omega_{\mathrm{K}}(t, p)}{\gamma_{\mathrm{K}}(p)}\right]\,.
\label{eq:sech}
\end{equation}
Here, $A_{\mathrm{K}}$ is the peak amplitude,  $\gamma_{\mathrm{K}}$ is the decay width of the peak and $\omega_{\mathrm{K}}$ is the corresponding dispersion relation, which can be extracted by employing \eqref{eq:sech} as a fit function.

Example data extracted from numerical simulations for $F(t,\omega,p)$ and $\rho(t,\omega,p)$ is shown in Fig.~\ref{fig:Ffun} for two dimensions, as functions of frequency $\omega-M$ relative to the `mass' gap of the dispersion, $\omega(p=0)=M$, for different times $t$ and different momenta $p$.
To obtain a clearer picture, it is useful to depict the statistical function $F(t,\omega,p)$ in the frequency-momentum plane, as shown in Fig.~\ref{fig:omega_p_resolved}, in both two (left) and three (right) dimensions, evaluated at time $t=1000$.  
As already denoted on these figures and explained in the following section, the $F_{\mathrm{K}}$ peak is identified with Kelvin waves and their quantised excitations, kelvons, in three dimensions, as well as the analogous kelvon quasiparticles in two dimensions.
From both representations, we can also see that at higher momenta, a second peak appears, which has been identified before as a Bogoliubov peak \cite{Boguslavski:2019ecc}. 
This Bogoliubov peak is more easily accessible from the spectral function $\rho(t,\omega,p)$, which is what we have used at $p=0$ to determine the effective mass gap $M$. 
The peak can be fitted by approximate Lorentzians of the form
\begin{equation}
\begin{aligned}
\rho(t, \omega, p) \simeq &\ \frac{2 A_{+}(p) \gamma_{+}(t, p)}{[\omega-\omega(p)]^2+\gamma_{+}(t, p)^2} \\
& -\frac{2 A_{-}(p) \gamma_{-}(t, p)}{[\omega+\omega(p)]^2+\gamma_{-}(t, p)^2},
\label{eq:bogfit}
\end{aligned}
\end{equation}
that capture the positive and negative branches of the Bogoliubov excitation. 
Moreover, its dispersion can be accurately fitted by using the relativistic generalisation \eq{BogRel} of the Bogoliubov dispersion \cite{Boguslavski:2019ecc}, which is shown in Fig.~\ref{fig:bogdisp}.

The two-dimensional $F_{\mathrm{K}}$ looks qualitatively very similar to the three-dimensional one, although our curves are noisier as statistical convergence needs considerably more simulation runs to average over in lower dimensions. 
Nevertheless, we are still able to accurately extract the peak properties in two dimensions from the available data.

We note that in general, each of the excitations in $F$ should have a counterpart in $\rho$. 
For the Bogoliubov peak, we confirm this for the two-dimensional theory in \Fig{Ffun} and refer to Ref.~\cite{Boguslavski:2019ecc} for a confirmation in $d=3$. 
In contrast, the identification of the $F_{\mathrm{K}}$ peak given by \eqref{eq:sech} in $\rho$ is challenging due to the peak's relative suppression in comparison to the Bogoliubov excitations. 
For $d=3$, the existence of such a peak in $\rho$ was demonstrated in Ref.~\cite{Boguslavski:2019ecc} by carefully analysing high-statistics data. 
For $d=2$, our level of statistics is insufficient to give a clear statement about its existence in $\rho$.
% ==================================================================== 
\subsection{Kelvon dispersion relations}
One of the most important aspects of distinguishing different kinds of quasiparticles and degrees of freedom is understanding their dispersion relation(s).
To this end, we have extracted the positions of the different peaks at multiple times.
As for the fit forms in the statistical function, up to $p<0.17$, we have used \eqref{eq:sech}, after which a two-peak fit is performed with one additional peak from Eq.~\eqref{eq:bogfit} that takes the Bogoliubov excitation into account in order to improve the accuracy of the fits.
Figure \ref{fig:combined_dispersion} shows the momentum-dependent position $\omega_\mathrm{K}(p)$ for the dominant infrared peak in (a) two and (b) three dimensions.

In three dimensions, the dispersion relation of this peak has been extracted before and was considered to be approximately linear \cite{PineiroOrioli:2018hst, Boguslavski:2019ecc}. 
However, upon closer inspection, one can see that it is, in fact, quadratic at low momenta and linear at higher momenta, which is characteristic for kelvons \cite{clark2015spatiotemporal}. 
Re-analysing the dispersion relation from Ref.~\cite{Boguslavski:2019ecc}, we find an excellent fit with the Kelvin wave dispersions as shown in Fig.~\ref{fig:combined_dispersion}b.
We fitted the long-wavelength limit of \Eq{kelvon}, with the additional zero-momentum shift proposed in Ref.~\cite{simula2008vortex}, 
\begin{equation}
\omega\left(p_0+p\right)=\omega_0+\frac{\Gamma p^2}{4 \pi} \ln \left(\frac{1}{\left|r_\mathrm{c} p\right|}\right), \quad\left|r_\mathrm{c} p\right| \ll 1\,,
\label{eq:kelvon2}
\end{equation}
where $\omega_0$ is the frequency of the kelvon with the smallest momentum $p_0\to0$ relative to the mass gap $M$ extracted from Bogoliubov excitations, $\Gamma$ is the (here positive) vortex circulation and $r_\mathrm{c}$ is the vortex core radius. 
These quantities show a time dependence, which represents characteristics of the scaling dynamics not extracted previously.
We find an increasing vortex core radius $r_c$, which reflects that in the course of the infrared transport, the number of coherent excitations at momenta on the order of the inverse healing length,\footnote{The healing length was extracted from the linear part of the Bogoliubov dispersion, where the speed of sound $c_\mathrm{s}$ and the healing length $p_\xi$ are related via $c_\mathrm{s}=p_\xi/\sqrt{2}M$.} $p_\xi=\xi_\text{h}^{-1}\approx 1.2\,Q$ in $d=2$ and $p_\xi\approx 0.6\,Q$ in $d=3$, decreases in time. 
This increase in the core radius as $r_c(t)\sim t^\beta$ is corroborated by the findings of \cite{Villois2016a.PhysRevE.93.061103}, where the total volume enclosed by the vortex cores in decaying Vinen turbulence in a superfluid in $d=3$ dimensions was found to settle to a constant at large times, while the vortex line density decays as $t^{-1}$.
The gap $\omega_0$, in $d=3$, is found to be negative and its modulus to decrease as $|\omega_0(t)|\sim t^{-2\beta}$, as can be observed in Fig.~\ref{fig:combined_dispersion}(b). 
This decrease is consistent with the expectation that $\omega_0$ corresponds to the energy of the Kelvin mode with the lowest wave number along the vortex lines, which should decrease as $p_0^2(t)\sim t^{-2\beta}$ for vortex rings with diameter increasing as $\ell_\text{v}(t)\sim t^\beta$.
As a result, the dispersions are found to obey the self-similar scaling relation 
\begin{equation}
\omega_{\mathrm{K}}(t, p)-M=t^{-\beta z} \tilde{\omega}_S\left(t^\beta p\right)\,,
\label{eq:dispselfs}
\end{equation} 
with exponent $\beta \approx 0.5$ and $z=2$ \cite{Boguslavski:2019ecc}, as also visible in the inset.
Here, $z=2$ reflects the dynamical scaling dimension of the Kelvin-wave dispersion relation.
At later times, we fit the dispersion \eqref{eq:kelvon2} across a progressively smaller momentum range, as shown by the dashed lines in Fig.~\ref{fig:combined_dispersion}(b).
This is necessary because the dispersion gradually starts to flatten at higher momenta, as clearly visible at $t=4000$, deviating from the expected form and making the fit increasingly unreliable. 
As a result, the extracted parameters become less meaningful in this regime. 
A more detailed discussion of this behaviour can be found in \App{kelvonflat}.

For two dimensions, we show the dispersion relation at different times in Fig.~\ref{fig:combined_dispersion}a. 
We find that it can be equally well fitted by both the kelvon and linear Bogoliubov dispersions, which we discuss more quantitatively in \App{fit} by investigating the residuals for both $d=2,3$ and both fits.
The dispersion obeys the self-similar scaling form \eqref{eq:dispselfs} as in $d=3$, but with exponents $\beta \approx 0.25$ and again $z \approx 2$, cf.~ inset of Fig.~\ref{fig:combined_dispersion}a.

%===========================================
\begin{figure*}[htb]
    \centering
    \begin{minipage}[t]{.47\textwidth}
        \centering
        \includegraphics[width=\textwidth]{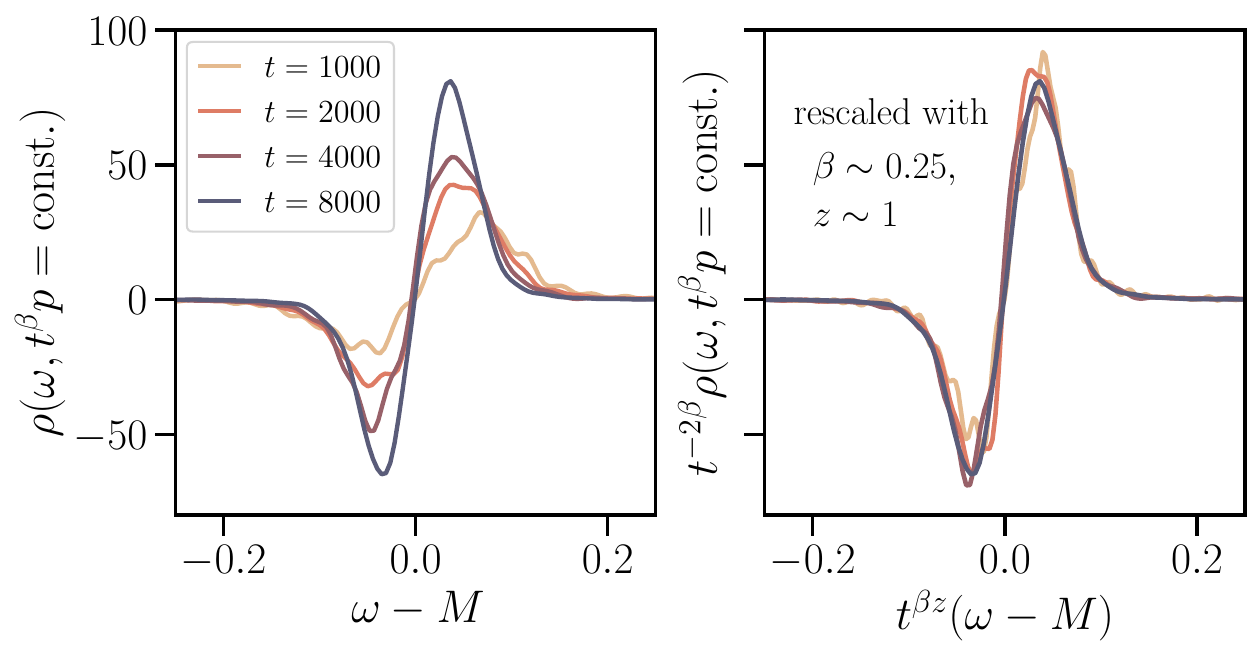}
        \caption{Self-similar evolution of the spectral function $\rho$ in $d=2$ showing the Bogoliubov peak (both positive and negative frequency branches) for constant slices of $t ^{\beta}p$ with $\beta \approx0.25$ at $p=0.21, 0.15, 0.13, 0.11$, for the respective times from $t=1000$ to $t=8000$.
        }
        \label{fig:spectralf}
    \end{minipage}
    \hfill
    \begin{minipage}[t]{.49\textwidth}
        \centering
        \includegraphics[width=\textwidth]{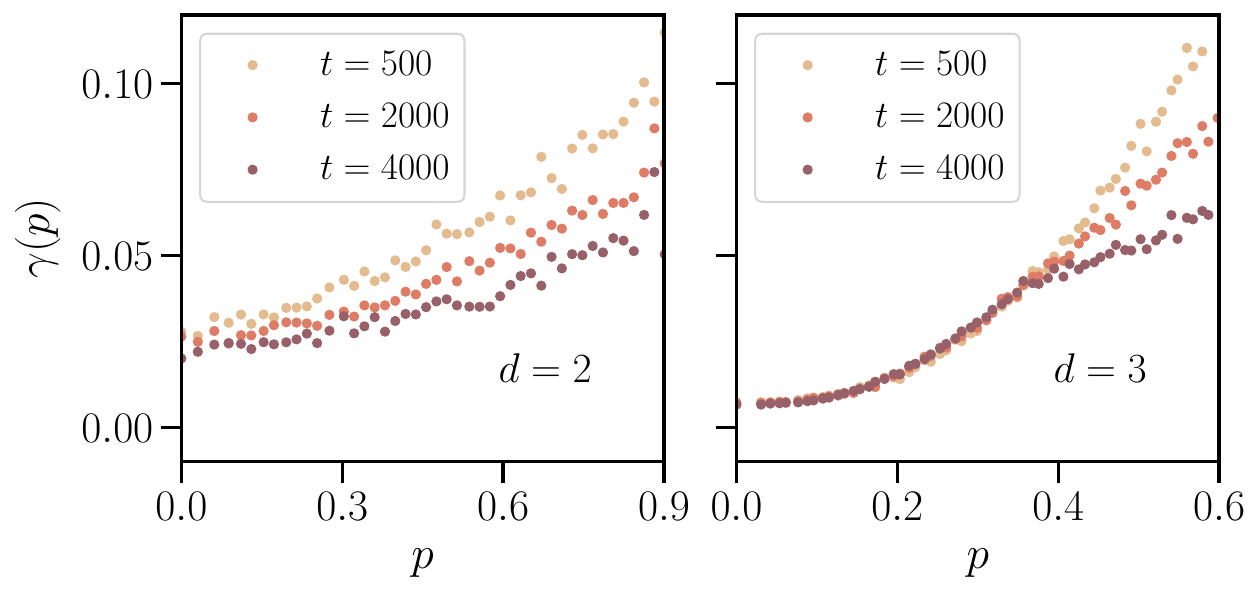}
        \caption{Time-dependent decay width of the kelvon peak for a single component scalar field theory in two (left) and three (right) dimensions.}
        \label{fig:width2D}
    \end{minipage}
\end{figure*}
% ===========================================

As already discussed earlier in \Sect{dof}, Kelvin waves along the vortex core are not possible in two dimensions. 
However, as long as one has a finite vortex core, kelvon-like quasiparticles can emerge and be parametrised, besides an azimuthal quantum number, by wave number $p$.
While the dispersion relation in Fig.~\ref{fig:combined_dispersion}(a) can be fitted equally well by both a linear curve and the kelvon dispersion \eqref{eq:kelvon2}, indicating the presence of phonons and/or kelvons, it looks qualitatively very similar to the dispersion in Fig.~\ref{fig:combined_dispersion}(b) which is more accurately described by the kelvon fit in $d=3$ dimensions. 
With this in mind, we suggest that the dominant infrared peak in two dimensions also originates from kelvons.
Apart from this, we also observe Bogoliubov phonons, which are excitations on top of the condensate and have larger dispersion values \eqref{eq:BogRel}. Therefore, the dominant Bogoliubov peak can be generally found to the right of the kelvon peak in \Fig{Ffun}. 

One of the prominent differences between two and three dimensions is that the peaks are considerably broader in two dimensions, as observed from the frequency-momentum resolved statistical function $F(t,\omega,p)$ in Fig.~\ref{fig:omega_p_resolved}. 
A similar effect occurs in non-Abelian gauge plasmas where the excitation spectrum in two dimensions exhibits a broad non-Lorentzian peak that remains broad even at late times \cite{Boguslavski:2021buh} while the quasiparticle peak in three-dimensional plasmas becomes narrower with time \cite{Boguslavski:2018beu}. 
In general, fluctuations are more significant in lower dimensions because of the reduced spatial volume available to average them out. 
This leads to more prominent noise and dissipation effects, which can then further broaden the peaks.

Additionally, the broadening in two dimensions might also reflect the stronger coupling between the different modes -- kelvons and phonons from vortex interactions -- due to the reduced dimensionality, where interactions and energy exchange between excitations are more pronounced. 
This interplay could obscure the distinction between different quasiparticle contributions, making it more challenging to isolate specific dispersion features as compared with the three-dimensional case.
While in three dimensions, a distinct quadratic bending at the lowest momenta can be observed in Fig.~\ref{fig:omega_p_resolved}b, extracting $F(t,\omega_{\mathrm{max}},p)$ as shown in Fig.~\ref{fig:combined_dispersion}b clarifies that, indeed, the dispersion is accurately described by a Kelvin-wave dispersion.
Moreover, this three-dimensional representation also shows a striking similarity to the ``space-time resolved mass spectrum'' in Ref.~\cite{clark2015spatiotemporal}, which used GPE simulations to study Kelvin and (Bogoliubov) sound waves. 

Curiously, kelvons seem to dominate the dynamics of the infrared peak in  $F(t,\omega,p)$, and we do not find any significant contribution from phonons arising, e.g., from vortex-ring annihilation. 
In two dimensions, this is likely the case as well, but the broader peaks make it more difficult to distinguish their contributions with certainty, leaving open the possibility that phonons may still play a role. 
How phonon generation occurs in vortex decay in two-dimensional superfluids is still a matter of active research, with different $n$-body processes put forward for the mechanisms \cite{Karl:2016wko, groszek2016onsager, yu2016theory, Baggaley2018a.PhysRevA.97.033601, kanai2024dynamical}.
We note that in our case, the process in two dimensions seems to be dominated by three-body collisions, where a vortex-antivortex pair and a lone vortex/antivortex are needed, as described in \Sect{Coarsening} and Refs.~\cite{Karl:2016wko, Deng:2018xsk}.

We conclude that, while the topological defects themselves do not directly show up in the excitation spectra of $F$ and $\rho$, excitations that arise on top of these defects do. 
Kelvons, quantised excitations of Kelvin waves, seem to be the dominant degree of freedom in the infrared regime of a three-dimensional single-component scalar field theory, while in two dimensions, kelvons, and potentially phonons that arise from the relaxation process of vortex-antivortex annihilation processes play a similar role. 

We finally remark that, despite the smooth dependence of the dispersions on momentum, as seen in Figs.~\fig{omega_p_resolved}--\fig{combined_dispersion}, as well as the correspondingly smooth occupation number spectra shown in \Fig{distribution}, which are dominated by the kelvon transport peak at low momenta, cf.~\Fig{Ffun}, the different Kelvon modes generically contribute to the overall field pattern in an incoherent manner.
Hence, the macroscopic occupations at low momenta give rise to a kind of quasicondensate, with similar occupancies spread over a wide range of momenta, but rather random relative phases between the modes, cf.~\cite{Mikheev:2018adp} for the case of sound modes. 
This is reminiscent of quasicondensates in one- and two-dimensional superfluids in equilibrium, at zero and non-zero temperatures, and gives rise to the chaotic displacement excitations of the vortices seen in \Fig{defectsmp2}.

% ======================================================================
\subsection{Self-similar scaling and decay width}
In Ref.~\cite{Boguslavski:2019ecc} it was shown that the transport/kelvon peak in both $F(t,\omega,p)$ and $\rho(t,\omega,p)$ exhibits self-similar scaling in the three-dimensional $\mathrm{O}(1)$ theory.
Similarly, in Ref.~\cite{PineiroOrioli:2018hst}, self-similar scaling was demonstrated for the three-dimensional $\mathrm{U}(1)$ theory for the same peak in $F(t,\omega,p)$ with the exponents $\beta \approx 1/2$ and $z=2$, and for the Bogoliubov peak in $\rho(t,\omega,p)$, with the same $\beta$ and $z \approx 1$.

A natural question is whether we find scaling in $F$ and $\rho$ for both peaks in the two-dimensional $\mathrm{O}(1)$ system. Let us start with the Bogoliubov peak. 
It is more easily accessed from the spectral function, as the kelvon peak in $\rho$ seems to be weak. 
The spectral function is shown in Fig.~\ref{fig:spectralf} for $d=2$ as a function of frequency at different times and momenta whose combination $t^{\,\beta} p$ is kept fixed. 
The right panel demonstrates that it follows the self-similar scaling ansatz 
\begin{equation}
\rho(t,\omega,p)=t^{\,\beta(2-\eta)} \rho_\mathrm{s}\left(t^{\,\beta z} (\omega-M), t^{\,\beta} p\right)\,,
\end{equation}
with $\eta \approx 0$.
While $z\approx1$, as expected for phonons, we find $\beta \approx 0.25$, just like for the kelvon peak, while $\beta \approx 0.5$ in constant slices of $t^{\,\beta} p$ failed to reproduce self-similarity.
These findings suggest that there might be an intrinsic connection between the Bogoliubov phonons and the kelvon peak.

In contrast to the Bogoliubov peak, we have found no evidence of a consistent self-similar scaling solution for the $F_{\mathrm{K}}$ peak of the two-dimensional $\mathrm{O}(1)$ theory. 
This is unexpected because the left panel of \Fig{combined_dispersion} displays self-similar scaling for the dispersion relation $\omega_{\mathrm{K}}(p)$. 
The problem lies rather in the decay width that scales appropriately in three dimensions but fails to scale in two dimensions.

The decay width of the kelvon peak as extracted from $F$ using the fit form \eqref{eq:sech} is shown in Fig.~\ref{fig:width2D} for two (left panel) and three (right) dimensions as a function of momentum at different times. 
In three dimensions, the decay width was found to be time-independent for low momenta and to have a quadratic form $\sim p^2$ \cite{Boguslavski:2019ecc}, as we confirm in the right panel.
In contrast, as shown in the left panel, it depends on time in a nontrivial way in two dimensions.
One finds a relatively large plateau, i.e., an approximately constant decay channel, in addition to a momentum-dependent decay width that decreases with time. 
This evolution spoils a simple scaling ansatz in $F(t, \omega, p)$. 

We finally note that in both $d=2$ and $d=3$ dimensions there is a finite decay width even at zero momentum. 
However, it is worth noting that in $d=2$, this offset seems to dominate $\gamma_{\mathrm{K}}(p)$ over a vast momentum region, while in $d=3$, the decay width is dominated by a quadratic form $\sim p^2$ already at low momenta, facilitating the self-similar scaling solution found in \cite{Boguslavski:2019ecc, PineiroOrioli:2018hst}.
% ==================================================================== 
% ==================================================================== 
\section{Discussion}
\label{sec:discussion}
We have investigated the universal nonequilibrium dynamics of single-component relativistic scalar field theories in $d=2,3$ dimensions.
Our results reveal a direct connection between the dominant excitations in the infrared regime of nonthermal fixed points and topological defects through Kelvin waves in three dimensions, and the analogous kelvon quasiparticles in two dimensions. 

In three dimensions, the dominant excitation has a dispersion relation with an excellent fit by the Kelvin wave dispersion from Eq.~\eqref{eq:kelvon}, while a similar dominant excitation in two dimensions has a dispersion, which can be similarly well fitted by either a linear, or a kelvon momentum dependence.
Both dispersion relations are time-dependent, which can indirectly result from the underlying vortex-antivortex annihilation processes and thus the temporal buildup of more kelvon excitations, leading to an effective increase of the vortex core size. 

A striking difference between two and three dimensions lies in the decay width of these excitations. 
In three dimensions, the kelvon peak exhibits a well-defined, time-independent width at low momenta, while in two dimensions, the peak is significantly broader, and the decay width does not scale in a simple manner in time.
This suggests that the interactions between kelvons, phonons, and vortices in two dimensions are more intricate, potentially due to stronger mode coupling and increased dissipation. 
The inability to cleanly separate phononic contributions from kelvon-like excitations in the spectral function in two dimensions raises further questions about the interplay between different degrees of freedom in driving the infrared transport.

Regarding the role of vortex coarsening in the observed scaling behaviour, the decay of the vortex (line) density follows a power-law scaling in both two and three dimensions, with exponents consistent with those of the scaling evolution of the momentum-dependent distribution function. 
This supports the interpretation that self-similar transport in these systems is closely tied to vortex dynamics.

The strong-wave-turbulence picture of the infrared dynamics of nonthermal fixed points is consistent with our results. However, the underlying degrees of freedom for the $N=1$ component scalar field theory are notably distinct from what is included in large-$N$ theories, which are Goldstone quasiparticles with quadratic dispersion.

Overall, our results complement existing large-$N$ approaches and shed light on the origins of universal scaling behaviour and the importance of vortex and coarsening dynamics in single-component scalar field systems in the context of nonthermal fixed points. 
By combining a detailed analysis of the topological defects with the underlying excitation spectrum and its dynamical properties, our work paves the way towards a more comprehensive understanding of universal scaling behaviour in $\mathrm{O}(N)$ scalar models.
In particular, this approach can also be applied to other low-$N$ component systems to obtain a consistent picture of the contributing quasiparticle excitations relevant for nonthermal universal scaling phenomena.
The results obtained here also mark an essential step in constructing a low-energy effective theory for the $N=1$ system, which needs to be built using the relevant degrees of freedom identified in this work.
%====================================================================
\begin{acknowledgments}
The authors thank J.~Berges, A.~N.~Mikheev, A.~Pi{\~n}eiro Orioli, D.~Proment, I.~Siovitz, and A.~Villois for discussions on the topics described here. 
This work is part of and funded by the Deutsche Forschungsgemeinschaft (DFG, German Research Foundation) under Germany's Excellence Strategy EXC 2181/1--390900948 (the Heidelberg STRUCTURES Excellence Cluster) and the Collaborative Research Centre, Project-ID No. 273811115, SFB 1225 ISOQUANT. 
This research was funded in part by the Austrian Science Fund (FWF) [10.55776/P34455].
The authors acknowledge support by the state of Baden-W{\"u}rttemberg through bwHPC (bwUniCluster 2.0), and by SFB 1225 ISOQUANT as well as Heidelberg University concerning the publication fee.
\end{acknowledgments}

% ==================================================================== 
% ==================================================================== 

\appendix
\begin{center}
\vspace*{2ex}\textbf{APPENDIX}\vspace*{-2ex}
\end{center}
% ==================================================================== 
% ==================================================================== 
\section{Vortex (line) density detection algorithm}
\label{app:vortdet}
\begin{figure}
    \centering
    \includegraphics[width=0.95\linewidth]{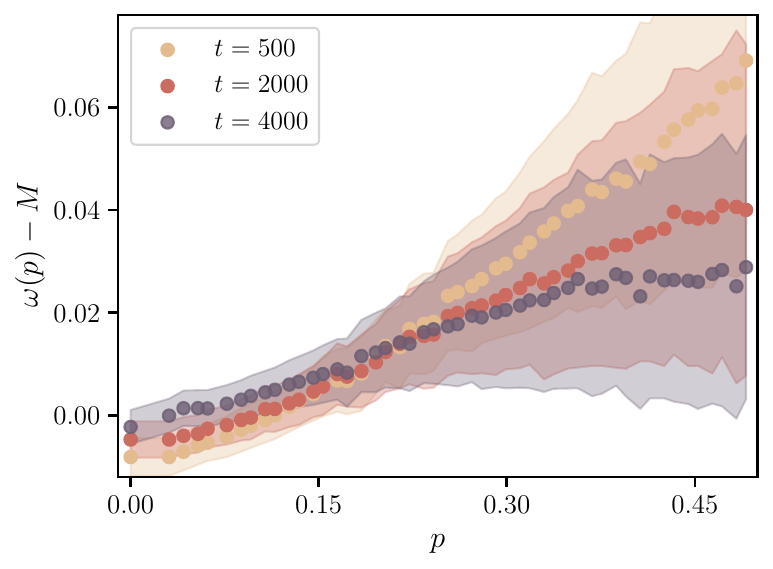}
    \caption{The Kelvin wave dispersion from Fig.~\ref{fig:combined_dispersion}(b), with the shaded area showing the width of the peaks in $F_{\mathrm{K}}(t,\omega,p).$}
    \label{fig:kelvonfitwidth}
\end{figure}
%===========================================
\begin{figure*}[htb]
    \centering
    \includegraphics[width=0.94\linewidth]{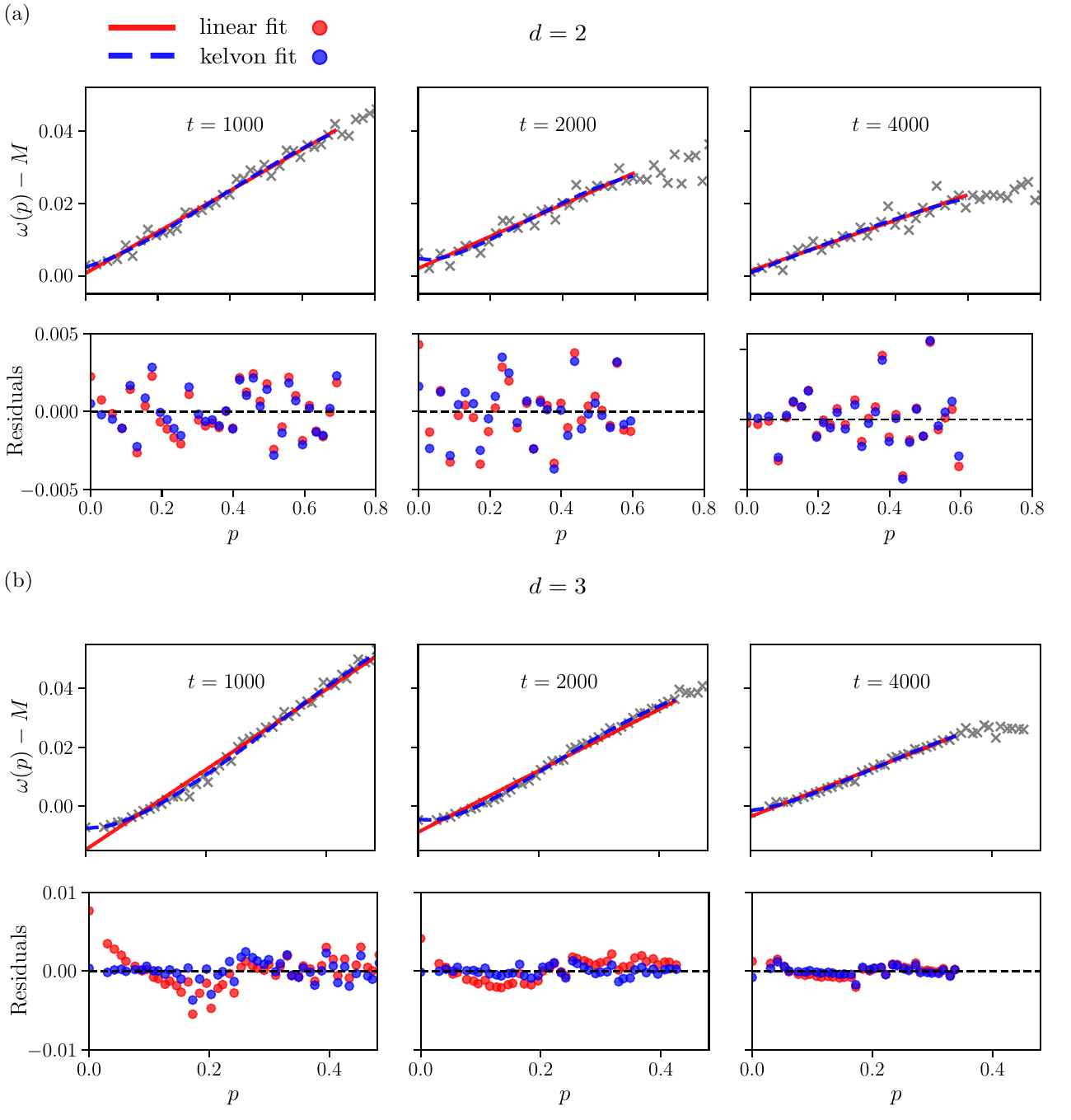}
    \caption{Comparison of linear and kelvon fits for the dispersion relation of the dominant infrared excitation in (a) $d=2$ and (b) $d=3$ dimensions, including the corresponding residuals. In $d=2$, it is challenging to determine which model provides a better fit. However, in $d=3$, the residuals strongly indicate that the kelvon fit captures the data more accurately.}
    \label{fig:fitres}
\end{figure*}
% ===========================================
Vortex defects are characterised by phase singularities in $\mathrm{arg}(\phi + \mathrm{i}\pi)$ where a sharp minimum in the energy density also occurs on the lattice.  
In two dimensions, we compute the vortex density by counting the total number of (anti)vortices in the phase $\mathrm{arg}(\phi + \mathrm{i}\pi)$, and divide this by the (constant) area of the system.
First, the input fields $\phi$ and $\pi$ are blockwise averaged over every set of $8\times8$ lattice points to reduce lattice artefacts. 
In two dimensions, this process reduces the original $2048^2$ lattice to a coarser $256^2$ grid.
For each point in the phase angle lattice, the algorithm then computes the phase differences around a closed loop. 
This loop consists of eight neighbouring points (excluding the edges of the lattice) arranged in a clockwise sequence, forming a square path centred around the point of interest. 
The phase differences between consecutive points in the loop are calculated by subtracting the phase angle at one point from the phase angle at the next point along the loop.
Eventually, these differences are summed up and if the total phase change around the loop is approximately $2\pi$, the point is identified as a vortex, and similarly, an antivortex is detected for a phase change of around $-2\pi$.
This procedure is designed to avoid mistakenly identifying points as defects where the phase fluctuates between $\pm \pi$, as such fluctuations do not result in a uniform phase winding around the point, and therefore do not indicate the presence of a vortex core.
Moreover, to avoid duplicate detections, which may still arise due to artefacts, the code enforces a minimum distance between detected vortices and antivortices.
After a parameter scan, this minimum distance was chosen as $5$ lattice points. 
This procedure is then repeated for each point on the $2048^2$ lattice.

However, phase singularities in a three-dimensional lattice are more challenging to compute, and doing the computation in the two-dimensional planes, plane by plane, could be misleading.
Therefore, we extract the vortex \emph{line} density \cite{Vinen1957a} from the energy density profiles, e.g.~the ones shown in \Fig{defectsmp2}.
We first blockwise average over every $6\times 6\times 6$ lattice points to reduce the original $368^3$ lattice to a coarser grid. 
At this point, the energy density becomes quite uniform, and the vortex tangles in \Fig{defectsmp2} are revealed by taking the points that are at $60\dots70\%$ of the mean energy value on the lattice. 
We then compute the number of points that belong to vortex lines, and divide it by the total number of points in the lattice, giving us the density of vortex lines. 

\section{Flattening of the kelvon dispersion curves}
\label{app:kelvonflat}
As mentioned in the main text, the Kelvin wave dispersion \eqref{eq:kelvon2} is fitted to the extracted dispersion relations over a progressively smaller momentum range at later times in three dimensions, since the fit becomes less reliable due to the flattening of the dispersion curves at higher momenta. 
This flattening is not part of the kelvon dispersion, therefore to reliably extract the fit parameters, we have to exclude this from the fit region. 
While the exact origin of this effect remains unclear, it is likely influenced by limitations in our data extraction procedure.
In the statistical function $F(t,\omega,p)$, at later times, and for higher momenta, the kelvon peak becomes weaker as the Bogoliubov peak starts to dominate. 
This makes the extraction of the position of the maximum of the peak challenging, which is the quantity that yields $\omega_{\mathrm{K}}$. 
Even though we start using a two-peak fit form after $p\geq0.17$, that includes both the kelvon \eqref{eq:sech} and the Bogoliubov peaks \eqref{eq:bogfit} in the fit function, the peak position of \eqref{eq:sech} becomes less certain. 
We also illustrate this by including the width $\gamma_{\mathrm{K}}$ from Fig.~\ref{fig:width2D} as a shaded region around $\omega_{\mathrm{K}}$ in Fig.~\ref{fig:kelvonfitwidth}.
While, naturally, the peak amplitude diminishes away from $\omega_{\mathrm{K}}$, as also visible in Fig.~\ref{fig:omega_p_resolved}, the increasing breadth of the peak indicates the growing uncertainty in the maximum position with increasing momentum.

\section{Analysing the fits to dispersion relation data}
\label{app:fit}
The comparison of linear and kelvon fits for the dispersion relation in $d=2$ and  $d=3$ dimensions reveals specific differences, as shown in \Fig{fitres}.
In $d=3$, the kelvon fit clearly captures both the expected $p^2$ behaviour at low $p$ and the bending of the data at higher $p$. 
This is reflected in the residuals, which are evenly distributed around zero for the kelvon fit, indicating that it accurately describes the data across the entire range. 
In contrast, the linear fit exhibits systematic deviations, as it fails to account for the $p^2$ scaling at low $p$  and the curvature at higher $p$, leading to structured residuals.

In $d=2$, however, the data is significantly more scattered, making it difficult to determine a clear preference between the two models. 
Both fits result in residuals that appear evenly distributed around zero, suggesting that the noise in the data obscures any systematic deviations. 
Consequently, while the kelvon fit is strongly favoured in $d=3$, the ambiguity in $d=2$ prevents a definitive conclusion about which model captures the underlying dispersion relation better.
However, since \Fig{omega_p_resolved} shows that in $d=2,3$ dimensions, the behaviour is qualitatively the same, it is strongly suspected that kelvon-like excitations also exist in $d=2$ dimensions.
\bibliographystyle{apsrev4-1}
%\newpage
\bibliography{new_file_clean}% Produces the bibliography via BibTeX.

\end{document}